\documentclass[pra,aps,showpacs,groupedaddress,superscriptaddress,twocolumn]{revtex4-1}

\usepackage[colorlinks=true,linkcolor=blue,urlcolor=blue,citecolor=blue]{hyperref}

\usepackage[utf8x]{inputenc}
\usepackage{color,soul}
\usepackage{bbm} 

\usepackage{amsfonts,amsmath,amssymb,stmaryrd}

\usepackage{graphicx}
\usepackage{subfigure}  % use for side-by-side figures
\usepackage{bbm} 
\usepackage{hyperref}
\usepackage{epsfig}
\usepackage{mathrsfs}
\usepackage{verbatim}
\usepackage{centernot}
\usepackage{ulem}

\renewcommand{\l}{\left(}
\renewcommand{\r}{\right)}

\newcommand{\bra}[1]{\langle#1|}
\newcommand{\ket}[1]{|#1\rangle}

\renewcommand{\H}{\hat{\mathcal{H}}}
\newcommand{\Ht}{\tilde{\mathcal{H}}}

\renewcommand{\a}{\hat{a}}

\newcommand{\ad}{\hat{a}^\dagger}

\newcommand{\h}[1]{\hat{#1}}

\newcommand{\G}{\hat{\Gamma}}

\newcommand{\hc}{\text{h.c.}}
\newcommand{\MF}{\text{MF}}

\newcommand{\f}{\text{F}}
\newcommand{\s}{\text{S}}
\renewcommand{\sf}{\text{MIX}}

\newcommand{\W}{\hat{W}}

\newcommand{\F}{\hat{F}}

\newcommand{\ph}{\text{ph}}
\newcommand{\IB}{\text{IB}}

\newcommand{\eff}{\text{eff}}

\usepackage{array}

\usepackage{cancel,ifthen}
\newcommand{\cmnt}[2][NoInPuT]{\ifthenelse{\equal{#1}{NoInPuT}}{}{{\color{red}\sout{#1}}} {\color{blue} #2}}

\usepackage{bm}	% bold in math mode
\renewcommand{\vec}[1]{\bm{#1}}

\begin{document}
\normalem	% changes \emph back to normal after introducing ulem package.

\title{Strong coupling Bose polarons out of equilibrium: Dynamical RG approach}

\author{F. Grusdt}
\affiliation{Department of Physics, Harvard University, Cambridge, Massachusetts 02138, USA}

\author{K. Seetharam}
\affiliation{Department of Electrical Engineering, Massachusetts Institute of Technologies, Cambridge, Massachusetts 02139, USA}
\affiliation{Department of Physics, Harvard University, Cambridge, Massachusetts 02138, USA}

\author{Y. Shchadilova}
\affiliation{Department of Physics, Harvard University, Cambridge, Massachusetts 02138, USA}

\author{E. Demler}
\affiliation{Department of Physics, Harvard University, Cambridge, Massachusetts 02138, USA}


\date{\today}

\begin{abstract}
When a mobile impurity interacts with a surrounding bath of bosons, it forms a polaron. Numerous methods have been developed to calculate how the energy and the effective mass of the polaron are renormalized by the medium for equilibrium situations. Here we address the much less studied non-equilibrium regime and investigate how polarons form dynamically in time. To this end, we develop a time-dependent renormalization group approach which allows calculations of all dynamical properties of the system and takes into account the effects of quantum fluctuations in the polaron cloud. We apply this method to calculate trajectories of polarons following a sudden quench of the impurity-boson interaction strength, revealing how the polaronic cloud around the impurity forms in time. Such trajectories provide additional information about the polaron's properties which are challenging to extract directly from the spectral function measured experimentally using ultracold atoms. At strong couplings, our calculations predict the appearance of trajectories where the impurity wavers back at intermediate times as a result of quantum fluctuations. Our method is applicable to a broader class of non-equilibrium problems. As a check, we also apply it to calculate the spectral function and find good agreement with experimental results. At very strong couplings, we predict that quantum fluctuations lead to the appearance of a dark continuum with strongly suppressed spectral weight at low energies. While our calculations start from an effective Fr\"ohlich Hamiltonian describing impurities in a three-dimensional Bose-Einstein condensate, we also calculate the effects of additional terms in the Hamiltonian beyond the Fr\"ohlich paradigm. We demonstrate that the main effect of these additional terms on the attractive side of a Feshbach resonance is to renormalize the coupling strength of the effective Fr\"ohlich model.
\end{abstract}

\maketitle

%%%%%%%%%%%%%%%%%%%%%%%%%%%%%%%%%%%%%%%%%%%%%%%%%%%%%
\section{Introduction}
\label{sec:Intro}
%%%%%%%%%%%%%%%%%%%%%%%%%%%%%%%%%%%%%%%%%%%%%%%%%%%%%

When a mobile impurity interacts with a surrounding medium, it becomes dressed by a cloud of excitations. In equilibrium, this leads to a renormalization of the impurity's properties such as its effective mass and energy. This effect can be understood more generally by the formation of a quasiparticle, the polaron, which is adiabatically connected to the free impurity \cite{Landau1946,Landau1948,Mahan2000}. The problem of how an impurity becomes modified by a surrounding medium has a long history \cite{Froehlich1954,deGennes1960}, and polarons have been observed in -- or near -- equilibrium in numerous systems \cite{PolaronsAdvMat2007,alexandrov2009advances,Hulea2006,Gershenson2006,Ortmann2011,Wang2015,Devreese2013}. They have also been realized recently using ultracold atoms \cite{Catani2012,Scelle2013,Fukuhara2013,Rentrop2016,Jorgensen2016PRL,Hu2016PRL}, where the tunability of inter-particle interactions \cite{Bloch2008,Chin2010} allows access to the strong coupling regime.

\begin{figure*}[t!]
\centering
\epsfig{file=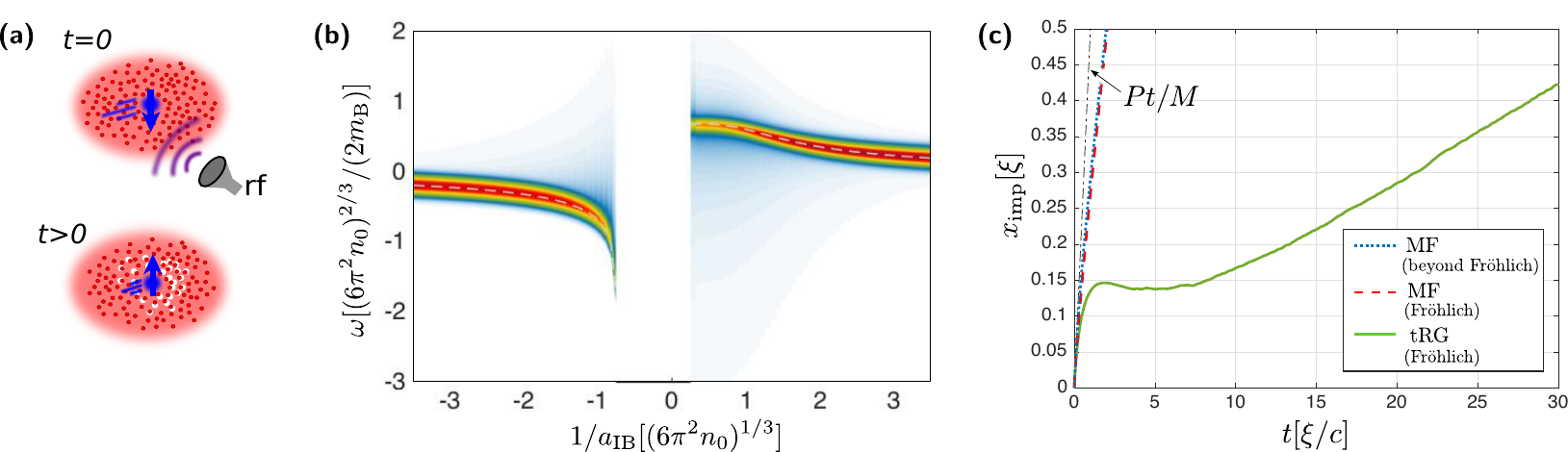, width=0.97\textwidth} $\quad$
\caption{(a) A mobile impurity atom in a non-interacting spin state can be transferred into an interacting spin state by a radio-frequency (RF) pulse. We describe the subsequent polaron dynamics using a time-dependent renormalization group approach (tRG). (b) When the RF pulse is weak, detecting the spin-flip probability allows measurement of the spectral function of the impurity. (c) When a strong RF $\pi$-pulse is used, interesting polaron formation dynamics can be observed. In this case, the trajectories of impurities at strong couplings and with a finite initial momentum can waver back at intermediate times. In (b), we used the Bogoliubov approximation but included two-phonon terms in the polaron Hamiltonian as in Refs.~\cite{Rath2013,Shchadilova2016PRL,Grusdt2017RG}. We used the same parameters as in the experiment by Hu et al. \cite{Hu2016PRL} with a mixture of $~^{40}{\rm K}$ impurities in a $~^{87}{\rm Rb}$ BEC: $n_0=1.8 \times 10^{14} {\rm cm}^{-3}$, $a_{\rm BB}=100 a_0$ and we employed a UV cut-off $\Lambda_0=10^3/\xi$ where $\xi$ is the corresponding healing length of the BEC. We performed calculations at vanishing polaron momentum, $P=0$. In (c), we solve the problem of an impurity with finite momentum $P>0$ (in the subsonic regime). The polaron trajectory $x_{\rm imp}(t)$ is shown after an interaction quench of the polaronic coupling constant from $\alpha=0$ to $\alpha=2.1$. The initial impurity velocity was $P/M=0.5 c$, and the mass ratio was $M/m_{\rm B}=0.5$. We used a sharp UV cut-off at $\Lambda_0=20/\xi$ in the calculations, and $c$ denotes the speed of sound in the BEC.}
\label{fig:Intro}
\end{figure*}

In this paper, we take the polaron problem to the next level and ask how a mobile impurity behaves in a far-from-equilibrium situation. More concretely, we consider a sudden quench of the interaction strength of the quantum impurity with a surrounding medium, varying the interaction strength from very weak to very strong values, see Fig.~\ref{fig:Intro} (a). We are interested in the subsequent dynamics on all time scales, ranging from short-time processes which can be treated perturbatively, intermediate scales where meta-stable pre-thermalized states can be reached, and long times where we investigate how the impurity equilibrates. 

Far-from-equilibrium situations, as described above, can be naturally realized in a well-controlled environment using experiments with ultracold atoms (see for example Refs.~\cite{Greiner2002,kinoshita2006quantum,hofferberth2007non,Hild2014}). In fact, the recent measurements of the polaron spectral function in the strong coupling regime \cite{Jorgensen2016PRL,Hu2016PRL} correspond to exactly this situation: strong impurity-boson interactions are suddenly switched on by flipping the spin of the impurity with the system's response subsequently recorded, see Fig.~\ref{fig:Intro} (b). As the quasiparticle weight of the polaron is strongly suppressed in this regime \cite{Shchadilova2016PRL,Grusdt2017RG}, the observed dynamics of the system involves states which vastly differ from the equilibrium polaron state. 

A second example concerns the trajectories of moving impurities; the trajectories can be imaged in a time-resolved manner after the quench \cite{Fukuhara2013,Catani2012}. This methodology has been utilized experimentally to measure the polaron's effective mass \cite{Catani2012} by investigating adiabatic polaron oscillations in a trapping potential \cite{Grusdt2016RG1D}. Here we consider a different aspect of this problem and calculate the impurity trajectory after a sudden interaction quench. Observing such trajectories allows study in real time of how the impurity slows down and a polaron forms while phonons are emitted, see Fig.~\ref{fig:Intro} (c).

%%%%%%%%%%%%%%%%%%%%%%%%%%%%%%%%%%%%%%%%%%%%%%%%%%%%%
%%%%%%%%%%%%%%%%%%%%%%%%%%%%%%%%%%%%%%%%%%%%%%%%%%%%%

In this paper, we investigate the effects of strong correlations and quantum fluctuations on the far-from-equilibrium dynamics of a mobile quantum impurity. Examining a many-body system with interacting bosons is a challenging task requiring powerful methods to unravel the system's physics. The restriction to a single impurity, however, makes the problem amenable to semi-analytical treatment, allowing us to gain important physical insight. To tackle this problem, we develop a semi-analytical time-dependent renormalization group (tRG) approach for solving non-equilibrium dynamics, taking into account quantum fluctuations in the polaron cloud. 

The study of far-from-equilibrium dynamics of quantum many-body systems is amongst the most challenging problems in theoretical physics. Many of the common approximations familiar from equilibrium problems can no longer be trusted or even applied. Nevertheless, progress has been made in solving the dynamics of several model systems. For example, in one dimension, the numerical DMRG method \cite{White1993,Schollwock2011} has been generalized for calculating time-dependent quantities and is now a widely used tool in these systems \cite{Vidal2004,White2004,Daley2004tDMRG}. In higher dimensions, the numerical DMFT method has also been generalized for solving dynamical problems \cite{Arrigoni2013}. Semi-analytical methods for solving dynamical problems, on the other hand, are much less developed. Time-dependent variational calculations, based on Dirac's variational principle \cite{Jackiw1980}, provide an important exception - the accuracy of their results, however, is not known in most cases. Another approach which can capture dynamics is Wegener's flow equation method \cite{Wegner1994} where the Hamiltonian is approximately diagonalized by a sequence of unitary transformations. Conceptually, our tRG approach is closely related to Wegener's method and is similar to the time-dependent RG approach introduced by Mathey and Polkovnikov in Ref.~\cite{Mathey2010}.

Our paper is organized as follows. After providing a brief overview of the research on Bose polarons in ultracold atoms in the remainder of this introduction, we discuss the formalism and introduce the model in Sec.~\ref{sec:Model}. Terms beyond the Fr\"ohlich paradigm are also included in the effective Hamiltonian. In Sec.~\ref{sec:Overview} we provide an overview of our method, sketch its basic principles, and present the tRG flow equations. The equations of motion are solved for concrete dynamical problems of mobile impurities in ultracold quantum gases, and we present our results in Sec.~\ref{sec:Results}. In Sec.~\ref{sec:tRG} full derivations of the tRG flow equations are provided. We conclude and give an outlook in Sec.~\ref{sec:Summary}.

%%%%%%%%%%%%%%%%%%%%%%
\subsection{Bose polarons in Bose-Einstein condensates}
\label{subsec:BosePolaronsBEC}
%%%%%%%%%%%%%%%%%%%%%% 

Bose polarons can be realized in a Bose-Einstein condensate (BEC) by introducing mobile impurity particles that interact with the particles of the host system \cite{Mathey2004,Tempere2009,Rath2013,Grusdt2015Varenna}. While the initial theoretical \cite{Mathey2004,Tempere2009,BeiBing2009,Casteels2011,Casteels2012,Shashi2014RF,Kain2014,Vlietinck2015,Grusdt2015RG,Shchadilova2016,Grusdt2016RG,Kain2016,Grusdt2016} and experimental \cite{Catani2012,Scelle2013,Fukuhara2013,Rentrop2016} work on this problem focused on the effective Fr\"ohlich model valid at weak couplings, it has been realized first in Ref.~\cite{Rath2013} that additional two-phonon terms need to be included in the Hamiltonian to provide an accurate description of Bose polarons at strong couplings \cite{Li2014,Levinsen2015,Christensen2015,Ardila2015,Shchadilova2016PRL,Parisi2017,Grusdt2017RG,Grusdt2016RG1D,Volosniev2017,Guenther2017,Yoshida2017,Levinsen2017,Sun2017,Sun2017a}. It has been argued recently that, on the attractive side of a Feshbach resonance, two-phonon terms mainly renormalize the effective Fr\"ohlich Hamiltonian describing the Bose polaron \cite{Grusdt2017RG}.

The experimental exploration of the strongly interacting regime has started recently with the measurement of the spectral function of the impurity \cite{Jorgensen2016PRL,Hu2016PRL}. The observed spectra are in good agreement with theoretical predictions which use truncated basis methods \cite{Rath2013,Li2014,Levinsen2015,Jorgensen2016PRL}, T-matrix calculations \cite{Rath2013} and time-dependent mean-field (MF) theory \cite{Shchadilova2016PRL}. This wide-spread agreement is somewhat surprising given how different the corresponding wavefunctions for the various approaches are expected to be \cite{Shchadilova2016PRL,Grusdt2017RG}. Indeed, to distinguish between different theoretical descriptions and understand the behavior of the polaron at strong couplings, additional measurements are required \cite{Grusdt2017RG}. For example, the direct observation of polaron dynamics of the impurity's trajectory after a sudden interaction quench provides a compelling alternative approach \cite{Catani2012,Grusdt2016RG1D}.

So far, theoretical work on Bose polaron dynamics has mostly focused on transport properties in one dimension \cite{Zvonarev2007,Bruderer2008,Johnson2011,Johnson2012,Schecter2012a,Schecter2012,Grusdt2014BO,Yin2015PRA,Grusdt2016RG1D}, and on the calculation of spectral functions \cite{Rath2013,Shashi2014RF,Shchadilova2016PRL,Jorgensen2016PRL}. Exceptions include analogies with Brownian motion \cite{Bonart2012,Lampo2017} and studies of trapped systems \cite{Bonart2013,Volosniev2015}. In this paper we develop a time-dependent RG approach to address general polaron problems far-from-equilibrium. To benchmark our method, we also calculate the spectral function, including the effects of quantum fluctuations and correlations between phonons. In Fig.~\ref{fig:Intro} (b), the result is shown for parameters relevant to the experiments in Ref.~\cite{Hu2016PRL} and using the Bogoliubov approximation for a weakly interacting BEC \cite{Rath2013,Shchadilova2016PRL,Grusdt2017RG}. 

Moreover, we derive tRG flow equations to calculate polaron trajectories following a sudden interaction quench, where the impurity starts with non-vanishing initial velocity below the speed of sound $c$ in the condensate. This problem has been addressed before using a time-dependent MF theory \cite{Shashi2014RF} with interesting dynamics predicted in the strong coupling regime. We find that interaction effects dramatically modify the polaron trajectory with phonon correlations leading to strong deviations from previous MF results. While the MF approach predicted damped oscillations \cite{Shashi2014RF}, our calculations show over-damped behavior; non-monotonic trajectories, however, can still be observed on intermediate time scales, see Fig.~\ref{fig:Intro} (c).

%%%%%%%%%%%%%%%%%%%%%%%%%%%%%%%%%%%%%%%%%%%%%%%%%%%%%
\section{Model}
\label{sec:Model}
%%%%%%%%%%%%%%%%%%%%%%%%%%%%%%%%%%%%%%%%%%%%%%%%%%%%%

The starting point for our analysis is the Bogoliubov-Fr\"ohlich Hamiltonian in $d=3$ dimensions ($\hbar = 1$),
\begin{equation}
\H =  \frac{\hat{\vec{p}}^2}{2 M}  + \int d^d \vec{k} ~ \Biggl[  V_{k} e^{i \vec{k}\cdot \hat{\vec{r}}} \l \a_{\vec{k}} + \ad_{-\vec{k}} \r  + \omega_{k} \ad_{\vec{k}} \a_{\vec{k}}  \Biggr].
\label{eq:HFroh}
\end{equation}
Here $\hat{\vec{p}}$ ($\hat{\vec{r}}$) denotes the impurity momentum (position) operator respectively, and $\a_{\vec{k}}$ annihilates a Bogoliubov phonon. The mass of the impurity is $M$. Within Bogoliubov theory the scattering amplitude is defined by
\begin{equation}
V_k = \sqrt{\alpha} \frac{c \sqrt{\xi}}{2 \pi \sqrt{2}} \l 1 + \frac{m_{\rm B}}{M} \r  \l \frac{k^2 \xi^2}{2 + k^2 \xi^2} \r^{1/4} 
\end{equation}
where $\alpha = a_{\rm IB}^2 / (a_{\rm BB} \xi)$ is the dimensionless coupling constant in 3d \cite{Tempere2009}, $a_{\IB}$ ($a_{\rm BB}$) is the impurity-boson (boson-boson) scattering length, $\xi$ and $c$ denote the healing length and the speed of sound in the BEC and $m_{\rm B} = 1 /( \sqrt{2} c \xi )$ is the mass of bosons in the BEC. The Bogoliubov dispersion is given by $\omega_k=c k \sqrt{1 + k^2 \xi^2/2}$. The dependence of $V_k$ and $\omega_k$ is specific for the BEC polaron, but the theoretical analysis below applies to any Hamiltonian of the type in Eq.~\eqref{eq:HFroh} in any spatial dimension $d$. 

The Hamiltonian \eqref{eq:HFroh} provides an accurate description of an impurity in a BEC when the mutual interactions between bosons and the impurity are weak, see, e.g., Refs.~\cite{Tempere2009,Grusdt2015Varenna} and Sec.~\ref{subsecBeyondFroehlich}. In the strong coupling regime, additional two-phonon terms need to be included \cite{Rath2013}; these terms lead to a renormalization of the effective Fr\"ohlich Hamiltonian \cite{Grusdt2017RG}. We discuss such corrections in Sec.~\ref{subsecBeyondFroehlich} and show for the attractive side of a Feshbach resonance how their main effect can be captured by using a renormalized coupling constant $\alpha^*$ in the Fr\"ohlich Hamiltonian. This result is established later in the paper by comparison of time-dependent MF calculations with and without the additional two-phonon terms.

The equilibrium properties of the Bogoliubov-Fr\"ohlich model have been discussed in detail in the literature using strong coupling theory \cite{Casteels2011}, weak-coupling MF theory \cite{BeiBing2009,Shashi2014RF}, perturbation theory \cite{Kain2014}, Feynman's path integral approach \cite{Tempere2009,Casteels2012,Vlietinck2015}, diagrammatic Monte Carlo techniques \cite{Vlietinck2015}, correlated Gaussian variational wavefunctions \cite{Shchadilova2016,Kain2016} and the RG method \cite{Grusdt2015RG,Grusdt2015Varenna,Grusdt2016RG}. In closely related works \cite{Bruderer2007,Bruderer2008a,blinova2013single,Rath2013,Li2014,Christensen2015,Levinsen2015,Ardila2015,Grusdt2017RG,Shchadilova2016PRL,Parisi2017,Grusdt2016RG1D,Volosniev2017} effects beyond the Fr\"ohlich model were also considered. 

Out of equilibrium, on the other hand, little is known about the system. In Refs.~\cite{Shashi2014RF,Grusdt2014BO,Shchadilova2016PRL} the time-dependent variational MF theory was used to calculate the spectral function as well as polaron dynamics in optical lattices. For strong couplings, dynamics have also been discussed using the adiabatic approximation \cite{Astrakharchik2004,Bruderer2008,Johnson2011,Johnson2012}. Modification of polaron dynamics in the intermediate coupling regime, however is poorly understood. Neither Feynman's variational all-coupling theory nor the diagrammatic Monte Carlo method can easily be generalized to the description of dynamics. In this paper we generalize the all-coupling RG method \cite{Grusdt2015RG,Grusdt2015Varenna,Grusdt2016RG} to non-equilibrium polaron problems and obtain the first results for polaron dynamics at intermediate couplings.

%%%%%%%%%%%%%%%%%%%%%%%%%%%%%%%%%%%%%%%%%%%%%%%%%%%%%
\subsection{Beyond the Fr\"ohlich paradigm}
\label{subsecBeyondFroehlich}
%%%%%%%%%%%%%%%%%%%%%%%%%%%%%%%%%%%%%%%%%%%%%%%%%%%%%
A more accurate description of an impurity in a BEC includes additional two-phonon terms going beyond the Fr\"ohlich model \cite{Rath2013}. They can be included in time-dependent MF calculations \cite{Shchadilova2016PRL}, allowing the estimation of their importance. On the attractive side of a Feshbach resonance, the main effect of two-phonon terms is to renormalize the effective Fr\"ohlich model describing the ground state of the model \cite{Grusdt2017RG}. In this regime, we now derive an expression for the renormalized coupling constant in the effective Fr\"ohlich model. 

Let us reconsider the Hamiltonian of an impurity at finite momentum $\vec{P}$ immersed in a weakly interacting BEC of ultracold atoms near a Feshbach resonance~\cite{Chin2010}. We make use of the Bogoliubov approximation and consider a weakly interacting bosonic gas with Bose-Bose contact interaction parameter $g_{\text{BB}}$. The approximation allows us to expand the bosonic system around the macroscopically occupied zero momentum state $\hat{\psi}_{\vec{k}=0}=\sqrt{n_{0}}$, where $n_0$ denotes the BEC density, and introduce Bogoliubov excitations around the condensate. 

The impurity interacts with the bosons locally, with the parameter of the contact interaction $g_{\Lambda_0}$ depending on the UV momentum cut-off $\Lambda_0$. As explained in Ref.~\cite{Shchadilova2016PRL}, this leads to the following microscopic Hamiltonian,
\begin{eqnarray}\label{eq:HbF}
&&\H  =\frac{\hat{\vec{p}}^{2}}{2M}+ \int d^d \vec{k} ~ \omega_{\vec{k}}\hat a_{\vec{k}}^{\dagger} \hat a_{\vec{k}} + g_{\Lambda_0} n_{0}  \\
&&+g_{\Lambda_0}\frac{\sqrt{n_{0}}}{(2\pi)^{3/2}} \int d^d \vec{k} ~ W_{\vec{k}}~ e^{i\vec{k}\vec{R}}\left(\hat a_{-\vec{k}}^{\dagger}+\hat a_{\vec{k}}\right)  \nonumber \\
&&+\frac{g_{\Lambda_0}}{(2\pi)^3} \int d^d \vec{k} d^d\vec{k'} ~V_{\vec{k},\vec{k}'}^{\left(1\right)}e^{i(\vec{k}-\vec{k}')\vec{R}} \hat a_{\vec{k}}^{\dagger} \hat a_{\vec{k}'}  \nonumber \\
 &&+\frac{g_{\Lambda_0}}{2 (2\pi)^3}\int d^d \vec{k} d^d\vec{k'} ~ V_{\vec{k},\vec{k}'}^{\left(2\right)}e^{i(\vec{k}-\vec{k}')\vec{R}} 
 \left( \hat a_{\vec{k}}^{\dagger}\hat a_{-\vec{k}'}^{\dagger}+\text{h.c.}\right)\nonumber 
\end{eqnarray}
where we define $W_{\vec k} = \left( \bigl( \frac{\vec{k}^2}{2m_{\rm B}} \bigr) / \bigl( \frac{\vec{k}^2}{2m_{\rm B}} + 2 g_{\rm BB} n_0 \bigr) \right)^{1/4}$, and two-particle interaction vertices are given by $V_{\vec k \vec{k'}}^{(1)} \pm V_{\vec k \vec{k'}}^{(2)} = \left( W_{\vec k}W_{\vec {k'}}\right)^{\pm 1/2}$. 

The first four terms of the Hamiltonian~\eqref{eq:HbF} constitute the Fr\"ohlich model~\eqref{eq:HFroh}. The relation between microscopic contact interaction strength $g_{\Lambda_0}$ and macroscopic scattering length $a_{\rm IB}$ differs between Fr\"ohlich model and the full Hamiltonian~\eqref{eq:HbF} which includes two-phonon scattering terms. The Fr\"ohlich model \eqref{eq:HFroh} with coupling constant $\alpha = a_{\rm IB}^2 / (a_{\rm BB} \xi)$ is obtained from the Born approximation result
\begin{equation}
g_{\Lambda_0}^{-1}= \frac{\mu_{\rm red}}{2\pi} a_{\rm IB}^{-1}
\end{equation}
where $\mu_{\rm red}=M m_{\rm B}/(M+m_{\rm B})$ is the reduced mass. 

The full Hamiltonian~\eqref{eq:HbF}, in contrast, allows for a proper regularization of the contact interaction between the impurity and bosons using the Lippmann-Schwinger equation, 
\begin{equation}\label{eq:LippSchwing}
g_{\Lambda_0}^{-1} =  \frac{\mu_{\rm red}}{2\pi} a_{\rm IB}^{-1} -\frac{1}{(2\pi)^3} \int^{\Lambda_0} d^d \vec{k} ~ \frac{2\mu_\text{red}}{\vec{k}^2}.
\end{equation}
Here the UV cut-off scale $\Lambda_0 \sim1/ r_0$ is related to a finite range $r_0$ of the impurity-boson interaction potential. In the limit $\Lambda_0 \to \infty$ contact interactions are recovered. 

The inclusion of the two-phonon scattering terms in~\eqref{eq:HbF} properly recovers the Feshbach resonance physics. Due to the interplay between the few-body physics generating the Feshbach resonance and the many-body enviornment, the position of the resonance is shifted as compared to the vacuum two-body scattering problem (impurity and single boson). This shift can be calculated analytically in the mean-field approximation for the zero total momentum case as \cite{Shchadilova2016PRL},   
\begin{equation}\label{eq:aSi}
a_{*}^{-1} = \frac{1}{(2\pi)^3 } \int^{\Lambda_0} d^d \vec{k} ~ \frac{\mu_{\rm red}}{2\pi}\left(\frac{2\mu_{\rm red}}{\vec{k}^2}-\frac{W_{\vec{k}}^2}{\omega_{\vec{k}}+\frac{\vec{k}^2}{2 M}}\right).
\end{equation}
We note that in principle the position of the resonance depends on the polaron momentum. However this dependence is weak and we will neglect it in this paper.

Comparing system dynamics under the Fr\"ohlich model to that under the full Hamiltonian~\eqref{eq:HbF} necessitates being the same relative distance from resonance in both cases. We can take the described resonance shift into account by introducing a renormalized effective scattering length for the Fr\"ohlich model,
\begin{equation}
[a_{\rm IB,Fr}(a_{\rm IB})]^{-1} = a_{\rm IB}^{-1} - a_{*}^{-1}.
\label{eqaIBFroehlichEff}
\end{equation}
The dimensionless interaction constant can thus be redefined as $\alpha^*(a_{\rm IB}) = [ a_{\rm IB,Fr}(a_{\rm IB})]^2 / (a_{\rm BB} \xi)$. In the weak-coupling regime, $a_{\rm IB} \to 0$, the Born approximation result is recovered, and at the shifted resonance $\alpha^* \to \infty$.

Using the mapping in Eq.~\eqref{eqaIBFroehlichEff}, we provide a direct comparison between the Fr\"ohlich and beyond-Fr\"ohlich models on a MF level in Sec.~\ref{sec:Results}. We emphasize, however, that this simple correspondence only applies on the attractive side of the Feshbach resonance ($a_{\rm IB,Fr}<0$), where the full Hamiltonian renormalizes to an effective Fr\"ohlich model \cite{Grusdt2017RG}. Additional bound states exist on the repulsive side, which are not captured by the Fr\"ohlich model but play a role in far-from-equilibrium dynamics \cite{Shchadilova2016PRL}.

%%%%%%%%%%%%%%%%%%%%%%
\subsection{Non-equilibrium problems}
\label{subsec:NonEqProb}
%%%%%%%%%%%%%%%%%%%%%%

We will now describe the specific non-equilibrium problems which we address in this paper using the tRG method and time-dependent MF theory. The approach is sufficiently general such that other problems can be considered as well, but for concreteness, we will restrict ourselves to two primary scenarios relevant for experiments with ultracold atoms. 

In both cases, we start from a non-interacting impurity at finite momentum $\vec{P}$. Phonons are assumed to be in their vacuum state $\ket{0}$ initially. Then, at $t=0$, the impurity-phonon interactions are suddenly switched on, and the system evolves coherently in time. Experimentally this scenario can be realized, for example, by quickly ramping the magnetic field close to a Feshbach resonance, or by flipping the internal state of the impurity from a non-interacting ($\downarrow$) to an interacting one ($\uparrow$). The second possibility is depicted in Fig.~\ref{fig:Intro} (a).

%%%%%%%%%%%%%%%%%%%%%%
\subsubsection{Dynamics of polaron formation}
\label{subsec:DynamicsOfPolaronFormation}
%%%%%%%%%%%%%%%%%%%%%%
The first question that naturally arises is how the trajectory of the impurity is modified when interactions are switched on \cite{Shashi2014RF}. We consider the case when the initial impurity momentum $P$ is sufficiently small, such that the emission of Cherenkov phonons is forbidden by conservation laws. The impurity will start to get dressed by phonons, forming a polaron, and correlations between the phonons begin to build up. As a consequence, the impurity slows down until a steady state is reached.

Classically one would expect that the impurity comes to complete rest at long times after its kinetic energy is emitted into phonons. However, quantum mechanically we obtain a steady state where the impurity is moving through the superfluid with a constant velocity. This can be understood by noting that equilibrium polaron ground states with non-zero momentum exist which sustain a finite impurity current, provided that the velocity is below the speed of sound \cite{Rath2013,Shashi2014RF}. As the initial state without phonons has a finite overlap with such equilibrium states, the average impurity momentum is non-vanishing in the steady state reached after the interaction quench.
 
When strong interactions are suddenly switched on, a large amount of energy is released into the system. Subsequently, this energy is divided between the polaron and the emitted phonons. We will show in Sec.~\ref{sec:Results} that this may result in polaron trajectories where the impurity wavers back at intermediate times before a steady state is reached, see for example Fig.~\ref{fig:Intro} (c).

Using the tRG method, we investigate how the impurity relaxes to a polaron at long times. The resulting steady state contains excitations in the form of emitted phonons on top of the true polaronic ground state, as can be seen from the conservation of energy after the interaction quench. We also investigate the properties of this steady state. The key observable to look at will be the time-dependence of the average impurity momentum, $\langle \hat{\vec{p}}(t) \rangle$. Using Ehrenfest's theorem, we can then calculate the trajectory of the impurity as
\begin{equation}
\langle \vec{x}(t) \rangle  = \int^t_0 d \tau ~ \frac{\langle \hat{\vec{p}}(\tau) \rangle}{M}.
\end{equation}

%%%%%%%%%%%%%%%%%%%%%%
\subsubsection{Spectral function}
\label{subsec:SpectralFunctionProblem}
%%%%%%%%%%%%%%%%%%%%%%
In a problem closely related to the interaction quench described above, one considers an impurity initialized in a non-interacting state $\downarrow$. By coupling it to an interacting state $\uparrow$, with a matrix element of strength $\Omega$ and frequency $\omega$, polarons can be created in the $\uparrow$ state. When the Rabi frequency $\Omega$ is sufficiently weak, this problem can be solved in linear response, and it follows that the probability for the impurity to be in the interacting $\uparrow$ state is proportional to the spectral function $I(\omega)$. 

Using Fermi's golden rule one obtains 
\begin{equation}
I(\omega) = \sum_n |\bra{\psi_\uparrow^n} \hat{S}_{\rm imp}^+ | \psi_\downarrow^0 \rangle |^2 \delta \l \omega - ( E_\uparrow^n - E_\downarrow^0 ) \r,
\label{eq:polaronSpectrum}
\end{equation}
where $\hat{S}_{\rm imp}^+ = | \! \uparrow \rangle \bra{\downarrow \! }$ describes a spin flip of the impurity; $\ket{\psi^0_\downarrow}$ denotes the ground state of the system at energy $E_\downarrow^0$ when the impurity is in its $\downarrow$ state and $\ket{\psi_\uparrow^n}$ are all eigenstates (labeled by $n$) at energies $E_\uparrow^n$ when the impurity is in its $\uparrow$ state. In the rest of this paper we consider only this so-obtained \emph{inverse RF spectrum}, as opposed to the \emph{direct RF spectrum} where an interacting state is flipped into a non-interacting one. 

For calculations of the spectral function we use a standard mapping to a dynamical problem. Eq.~\eqref{eq:polaronSpectrum} can be recast in the form
\begin{equation}
I(\omega) = \text{Re} \frac{1}{\pi} \int_0^\infty dt ~ e^{i \omega t} A(t),
\label{eq:IFTAt}
\end{equation}
see e.g. Ref.~\cite{Shashi2014RF}. Here the time-dependent overlap (related to the Loschmidt-echo, see e.g. \cite{Silva2008}) is defined as
\begin{equation}
A(t) = e^{i E_\downarrow^0 t} \bra{0} e^{- i \H t} \ket{0}.
\label{eq:defAt}
\end{equation}
It describes the amplitude for the phonons to return to their initial vacuum state $\ket{0}$ after the system has evolved in time, $\ket{0} \to e^{- i \H t} \ket{0}$, while the impurity is interacting with the phonons. This problem is closely related to the problem of polaron formation. Below we will use the tRG to calculate the time-dependent overlap. Unlike usual physical observables (e.g. the phonon momentum), the time evolution contains only the forward direction. Thus, the time-dependent overlap requires a special treatment.

%%%%%%%%%%%%%%%%%%%%%%%%%%%%%%%%%%%%%%%%%%%%%%%%%%%%%
\section{Overview of the method}
\label{sec:Overview}
%%%%%%%%%%%%%%%%%%%%%%%%%%%%%%%%%%%%%%%%%%%%%%%%%%%%%

Before we start to develop the tRG method for the Fr\"ohlich Hamiltonian \eqref{eq:HFroh}, we perform the same steps as in the equilibrium RG \cite{Grusdt2015RG,Grusdt2015Varenna,Grusdt2016RG} and bring the Hamiltonian into a more convenient form. To this end we first apply the unitary polaron transformation introduced by Lee, Low and Pines (LLP) \cite{Lee1953},
\begin{equation}
\h{U}_\text{LLP} = e^{i \h{S}} ,\qquad \h{S} = \h{\vec{r}} \cdot \h{\vec{P}}_{\rm ph},
\label{eq:LLPdef}
\end{equation}
where the total phonon momentum is given by $\h{\vec{P}}_{\rm ph} = \int d^dk ~ \vec{k}  ~\ad_{\vec{k}} \a_{\vec{k}}$. In the new frame the impurity is localized in the origin and the resulting Hamiltonian,
\begin{multline}
\H_{\vec{P}} =  \frac{1}{2 M} \l \vec{P} - \int d^d \vec{k} ~ \vec{k} \ad_{\vec{k}} \a_{\vec{k}} \r^2 \\
+ \int d^d \vec{k} ~ \Biggl[  V_{k}  \l \a_{\vec{k}} + \ad_{-\vec{k}} \r  + \omega_{k} \ad_{\vec{k}} \a_{\vec{k}}  \Biggr],
\label{eq:HFrohLLP}
\end{multline}
commutes with the momentum operator $\h{\vec{p}} = \vec{P}$ which takes the role of the conserved total momentum $\vec{P}$ of the system, see Refs.~\cite{Devreese2013,Grusdt2015Varenna} for review. In the following discussion we will always assume a given value $\vec{P}$ of the total momentum.  

Next we change into the frame of quantum fluctuations around the MF solution $\alpha^\MF_{\vec{k}}$ by applying the unitary MF shift
\begin{equation}
\hat{U}_{\rm MF} = \exp \l \int d^d \vec{k} ~ \alpha^{\rm MF}_{\vec{k}} ~ \ad_{\vec{k}} - \hc \r.
\label{eq:UMF}
\end{equation}
The MF amplitude is given by \cite{Shashi2014RF} $\alpha_{\vec{k}}^\MF = - V_k / \Omega_{\vec{k}}^\MF$ where the phonon dispersion in the new frame is
\begin{equation}
\Omega^\MF_{\vec{k}} = \omega_k + \frac{k^2}{2 M} - \frac{1}{M } \vec{k} \cdot \l \vec{P} - \vec{P}_\ph^\MF \r.
\label{eq:OmegakDef}
\end{equation}
Here $\vec{P}_\ph^\MF$ denotes the MF phonon momentum,
\begin{equation}
\vec{P}_\ph^\MF = \int d^d\vec{k} ~ \vec{k} |\alpha^\MF_{\vec{k}}|^2.
\label{eq:XiselfCons}
\end{equation}
The last expression defines the self-consistency equation of Lee-Low-Pines MF theory.

The Hamiltonian $\tilde{\mathcal{H}} = \h{U}^\dagger_\MF  \h{U}^\dagger_{\rm LLP} \H \h{U}_{\rm LLP} \h{U}_{\rm MF} $ in the new frame can be written in a very compact form now \cite{Grusdt2015RG}. Using generalized notations which will become useful later in the RG, we obtain for fixed $\vec{P}$
\begin{multline}
\tilde{\mathcal{H}}= E_0 + \int^{\Lambda} d^d \vec{k} ~ \ad_{\vec{k}} \a_{\vec{k}} \Omega_{\vec{k}}  \\
+ \int^{\Lambda} d^d \vec{k} ~ d^d \vec{k}' ~ \frac{1}{2}  k_\mu \mathcal{M}_{\mu \nu}^{-1} k_\nu'  ~ : \G_{\vec{k}} \G_{\vec{k}'} :,
\label{eq:HquantFluc2}
\end{multline}
where $\mu, \nu=x,y,...$ denote spatial coordinates (which are summed over according to Einstein's convention) and $:...:$ stands for normal ordering. We have introduced an ultra-violet (UV) momentum cut-off $\Lambda$ at high energies for regularization and defined operators 
\begin{equation}
\G_{\vec{k}}(\Lambda) = \alpha_{\vec{k}}(\Lambda) ( \a_{\vec{k}} + \ad_{\vec{k}} ) + \ad_{\vec{k}} \a_{\vec{k}}.
\end{equation}
The phonon dispersion in the new frame reads
\begin{equation}
\Omega_{\vec{k}}(\Lambda) = \omega_k + \frac{1}{2} k_\mu \mathcal{M}_{\mu \nu}^{-1}(\Lambda) k_\nu + k_\mu \mathcal{M}^{-1}_{\mu \nu}(\Lambda) \kappa_\nu(\Lambda)
\label{eq:OmegaDefRG}
\end{equation}
and both coupling constants $\mathcal{M}_{\mu \nu}(\Lambda)$ and $\kappa_\nu(\Lambda)$ will be flowing in the tRG. The coherent amplitude is given by $\alpha_{\vec{k}}(\Lambda) = - V_k / \Omega_{\vec{k}}(\Lambda)$, similar to the MF expression. Note that this leads to a dependence of the operators $\G_{\vec{k}}(\Lambda)$ on the UV cut-off $\Lambda$.

Before applying the RG protocol to effectively eliminate high-energy phonons from the problem, the UV cut-off is set to a constant $\Lambda=\Lambda_0$. This is where the initial conditions for the tRG protocol are defined,
\begin{equation}
\kappa_\mu(\Lambda_0) = \delta_{\mu x} \l P_{\rm ph}^{\rm MF} - P \r, \quad \mathcal{M}_{\mu \nu}(\Lambda_0) = \delta_{\mu \nu} M.
\end{equation}
In the first expression we assumed for simplicity that the total system momentum always points along the $x$-direction, i.e. $\vec{P} = P \vec{e}_x$. Note that the coherent amplitudes start from the MF result, $\alpha_{\vec{k}}(\Lambda_0) = \alpha_{\vec{k}}^\MF$.

%%%%%%%%%%%%%%%%%%%%%%
\subsection{tRG method -- physical observables}
\label{subsec:tRGOverview}
%%%%%%%%%%%%%%%%%%%%%%
One of the goals of this paper is to calculate the dynamics of physical observables $\hat{O}$ in the polaron problem, defined in the lab frame by
\begin{equation}
O(t) = \bra{\psi_0} e^{i \H t} \hat{O} e^{- i \H t} \ket{\psi_0},
\end{equation}
where $\ket{\psi_0}$ is the initial state. For simplicity we restrict ourselves to observables which do not involve correlations between different phonon momenta in the polaron frame and  hence can be written as
\begin{equation}
  \hat{O} = \hat{U}_{\rm LLP} \int^{\Lambda_0} d^d \vec{k} ~ \hat{O}_{\vec{k}} ~ \hat{U}_{\rm LLP}^\dagger.
\end{equation}
Here we assume that operators $\hat{O}_{\vec{k}}$ involve only phonons $\a_{\vec{k}}$, $\ad_{\vec{k}}$ at momenta $\vec{k}$.

Now we outline the basic structure of the tRG approach for the calculation of time-dependent observables $O(t)$. Similar ideas can be applied to the calculation of the time-dependent overlap, see Eq.~\eqref{eq:defAt}, although in that case the Hamiltonian generated in the RG flow can become non-Hermitian because the amplitude $A(t) \in \mathbb{C}$ is complex-valued in general and only one time-direction is involved. All details can be found in Sec.~\ref{sec:tRG}.

As a first step, we formulate the problem in the frame of quantum fluctuations around the MF polaron, i.e. we introduce the unitary transformation $\hat{U}_\MF$ to obtain
\begin{equation}
O(t) = \int^{\Lambda_0} d^d \vec{k} ~ \bra{\tilde{\psi}_0}  e^{ i \tilde{\mathcal{H}} t} \hat{o}_{\vec{k}} e^{- i \tilde{\mathcal{H}} t} \ket{\tilde{\psi}_0}.
\label{eq:OtFlucFrame}
\end{equation}
Here we have defined $\hat{o}_{\vec{k}} = \hat{U}^\dagger_\MF  \hat{O}_{\vec{k}} \hat{U}_\MF$ and the initial state in the polaron frame reads $\ket{\tilde{\psi}_0} = \hat{U}_{\rm MF}^\dagger \hat{U}_{\rm LLP}^\dagger \ket{\psi_0}$.

The key idea of the tRG method is to introduce another set of unitary transformations $\hat{U}_{\Lambda}$ in Eq.~\eqref{eq:OtFlucFrame}. They are chosen such that the Hamiltonian is diagonalized for fast phonon degrees of freedom in a small shell with momenta $\vec{k}$ between $\Lambda - \delta \Lambda < |\vec{k}| \leq \Lambda$, where $\delta \Lambda$ can be infinitesimally small. Repeating this procedure shell by shell reduces the UV cut-off from $\Lambda_0$ ultimately down to zero. Conceptually this approach is similar to the procedure used in the equilibrium RG \cite{Grusdt2015RG,Grusdt2015Varenna,Grusdt2016RG} for finding the ground state. The main difference is that out-of-equilibrium fast phonons can be in excited states, modifying the effect on slow phonons during the RG procedure. 

We start by applying the unitary transformations $\hat{U}_{\Lambda}$ to the Hamiltonian,
\begin{equation}
\hat{U}^\dagger_{\Lambda} \tilde{\mathcal{H}}(\Lambda) \hat{U}_\Lambda =  \tilde{\mathcal{H}}^{(0)}(\Lambda-\delta \Lambda) + \int_\f d^d \vec{k} ~ \ad_{\vec{k}} \a_{\vec{k}} \l \Omega_{\vec{k}} + \hat{\Omega}_\s(\vec{k}) \r.
\label{eq:tRGhamiltonianRenormalization}
\end{equation} 
Here $\tilde{\mathcal{H}}^{(0)}(\Lambda-\delta \Lambda)$ is a renormalized Hamiltonian of the form \eqref{eq:HquantFluc2} involving slow phonons with momenta $|\vec{p}| \leq \Lambda - \delta \Lambda$ only (we label slow phonons by $\s$). The right-most term describes dynamics of fast phonons (labeled by $\f$) with momenta $\Lambda -\delta \Lambda < |\vec{k}| \leq \Lambda$. The frequency of fast phonons is modified by a term $\Omega_\s(\vec{k})$ which involves only slow-phonon operators $\a_{\vec{p}}$, $\ad_{\vec{p}}$. 

Assuming that the frequency renormalization is small, $|| \hat{\Omega}_\s(\vec{k}) || \ll \Omega_{\vec{k}}$, the last term in Eq.~\eqref{eq:tRGhamiltonianRenormalization} can be treated perturbatively. To leading order the frequency renormalization $\hat{\Omega}_{\rm S}(\vec{k})$ has no effect on the fast phonon dynamics, which is then determined only by $\tilde{\mathcal{H}}_\f = \int_\f d^d \vec{k} ~ \ad_{\vec{k}} \a_{\vec{k}} \Omega_{\vec{k}}$. We thus obtain additional renormalization of the slow-phonon Hamiltonian,
\begin{equation}
\delta \tilde{\mathcal{H}}_\s(t) = \int_\f d^d \vec{k} ~ ~_\f \bra{\tilde{\psi}_0} e^{ i \tilde{\mathcal{H}}_\f t} \ad_{\vec{k}} \a_{\vec{k}}  e^{- i \tilde{\mathcal{H}}_\f t} \ket{\tilde{\psi}_0}_\f ~ \hat{\Omega}_\s(\vec{k}),
\label{eq:deltaHsDyn}
\end{equation}
where we assumed for simplicity that the initial state factorizes into contributions from fast and slow phonons respectively after the RG step,
\begin{equation}
\hat{U}_{\Lambda}^\dagger \ket{\tilde{\psi_0}} = \ket{\tilde{\psi_0}}_\s \otimes \ket{\tilde{\psi_0}}_\f.
\end{equation}

In the equilibrium RG, only the renormalization described by $\tilde{\mathcal{H}}^{(0)}(\Lambda-\delta \Lambda)$ was relevant. Out of equilibrium we obtain the additional term $\delta \tilde{\mathcal{H}}_\s(t)$ and the new Hamiltonian describing slow phonons reads
\begin{equation}
\tilde{\mathcal{H}}(\Lambda - \delta \Lambda) = \tilde{\mathcal{H}}^{(0)}(\Lambda-\delta \Lambda) + \delta \tilde{\mathcal{H}}_\s(t).
\label{eq:HslowRenormalized}
\end{equation}
In general, this Hamiltonian can depend on time $t$ explicitly, but in the Fr\"ohlich problem $\tilde{\mathcal{H}}_\f$ conserves the phonon number and thus Eq.~\eqref{eq:deltaHsDyn} is time-independent. By comparing the new Hamiltonian \eqref{eq:HslowRenormalized} to the universal expression \eqref{eq:HquantFluc2}, the tRG flow equations for the coupling constants can be derived. Our detailed calculations for the Fr\"ohlich model will be presented in Sec.~\ref{sec:tRG}. 

Now we return to the observables of interest, Eq.~\eqref{eq:OtFlucFrame}. By introducing unitaries $\hat{U}_\Lambda$ we will show in Sec.~\ref{sec:tRG} that an expression of the following form is obtained:
\begin{multline}
O(t) = \int_\f d^d \vec{k} ~ ~_\f \bra{\tilde{\psi}_0}  e^{ i \tilde{\mathcal{H}}_\f t} \hat{U}^\dagger_\Lambda \hat{o}_{\vec{k}} \hat{U}_\Lambda e^{- i \tilde{\mathcal{H}}_\f t} \ket{\tilde{\psi}_0}_\f \\
+ \int_\s d^d \vec{p} ~ ~_\s \bra{\tilde{\psi}_0}  e^{ i \tilde{\mathcal{H}}(\Lambda-\delta \Lambda) t} \hat{o}_{\vec{p}} e^{- i \tilde{\mathcal{H}}(\Lambda-\delta \Lambda) t} \ket{\tilde{\psi}_0}_\s. 
\label{eq:OtFlucFrameRen}
\end{multline}
The term in the first line describes the contribution of fast phonons to the observable $O(t)$, which can be cast in the form of a tRG flow equation for $O(t;\Lambda)$. The final expression is obtained when the limit $\Lambda \to 0$ is performed. Note that the time-dependence in this expression is purely harmonic and can thus be calculated analytically. The second line of Eq.~\eqref{eq:OtFlucFrameRen} describes slow phonon contributions and has a similar form as Eq.~\eqref{eq:OtFlucFrame}, which was the starting point of our analysis. The tRG can be applied to this expression again, and in this way, a tRG flow is generated.

%%%%%%%%%%%%%%%%%%%%%%
\subsection{tRG flow equations -- physical observables}
%%%%%%%%%%%%%%%%%%%%%%
The calculations described in the last paragraph are somewhat cumbersome, so we postpone their detailed discussion to Sec.~\ref{sec:tRG}. Here we summarize the tRG flow equations for the phonon number and momentum, which will be solved numerically in the following section. 

For the renormalized impurity mass we obtain the following RG flow equation,
\begin{equation}
\frac{\partial \mathcal{M}_{\mu \nu}^{-1} }{\partial \Lambda} = 2 \mathcal{M}_{\mu \lambda}^{-1}  \int_\f d^{d-1} \vec{k} ~ \frac{V_k^2}{\Omega^3_{\vec{k}}} k_\lambda k_\sigma  ~\mathcal{M}_{\sigma \nu}^{-1}, \vspace{0.1cm}
\label{eq:RGflowMass} 
\end{equation}
where $\int_\f d^{d-1} \vec{k}$ denotes the integral over the $(d-1)$-dimensional momentum shell with radius $\Lambda$. For the momentum $\kappa_x$ (recall that $\vec{P}=P \vec{e}_x$) we derive
%\begin{widetext}
%\begin{equation}
%\frac{\partial \kappa_x }{\partial \Lambda} =- \frac{\partial \mathcal{M}_{xx}^{-1}}{\partial \Lambda} \mathcal{M}_{xx} \kappa_x +  \l 1 + 2 \mathcal{M}_{xx}^{-1} I^{(2)} \r^{-1} %\\ \times
%\left[ 2 \mathcal{M}_{xx}^{-1} I^{(2)}  \l \int_\f d^{d-1}\vec{k} ~ k_x |\lambda_{\vec{k}}(t)|^2 \r  - I^{(3)}_{\mu \nu} \frac{\partial \mathcal{M}_{\mu \nu}^{-1} }{\partial \Lambda}   \right] .
%\label{eq:MFRGflowkappaX}
%\end{equation}
%\end{widetext}
\begin{multline}
\frac{\partial \kappa_x }{\partial \Lambda} =- \frac{\partial \mathcal{M}_{xx}^{-1}}{\partial \Lambda} \mathcal{M}_{xx} \kappa_x +  \l 1 + 2 \mathcal{M}_{xx}^{-1} I^{(2)} \r^{-1} \\ \times
\left[ 2 \mathcal{M}_{xx}^{-1} I^{(2)}  \l \int_\f d^{d-1}\vec{k} ~ k_x |\lambda_{\vec{k}}(t)|^2 \r  - I^{(3)}_{\mu \nu} \frac{\partial \mathcal{M}_{\mu \nu}^{-1} }{\partial \Lambda}   \right] .
\label{eq:MFRGflowkappaX}
\end{multline}
Here we introduced the following integrals,
\begin{flalign}
I^{(2)}(\Lambda) &= \int^\Lambda d^d \vec{k} ~ k_x^2 \frac{V_k^2}{\Omega^3_{\vec{k}}}, \label{eq:I2Def} \\
 I^{(3)}_{\mu \nu} (\Lambda) &=  \int^\Lambda d^d \vec{k} ~ k_x k_\mu k_\nu \frac{V_k^2}{\Omega^3_{\vec{k}}}, \label{eq:I3munuDef}
\end{flalign}
and we define
\begin{equation}
\lambda_{\vec{k}}(t) = -\alpha_{\vec{k}} \left[ 1 + \frac{1}{\Omega_{\vec{k}}}  k_\mu \mathcal{M}_{\mu \nu}^{-1} \int_\s d^d \vec{p} ~  p_\nu  |\alpha_{\vec{p}}|^2   \right] e^{- i \Omega_{\vec{k}}t}.
\end{equation}
For the zero-point energy $E_0$, which is given initially by the MF ground state energy $E_0(\Lambda_0) = E_0 |_{\rm MF}$, we obtain
\begin{equation}
\frac{\partial E_0}{\partial \Lambda} = \frac{1}{2} \frac{\partial \mathcal{M}_{\mu \nu}^{-1}}{\partial \Lambda} \int_\s d^d \vec{p} ~ p_\mu p_\nu \l \alpha_{\vec{p}} \r^2.
\label{eq:RGflowZeroPtEnergy}
\end{equation}

The tRG flow equation for the phonon momentum $P_\ph(t) = \lim_{\Lambda \to 0} P_\ph(t,\Lambda)$ contains an auxiliary variable $\chi(t,\Lambda)$ which is also flowing in the tRG. Its origin will become clear later; at this stage it is merely required to calculate the tRG flow of $P_\ph(t,\Lambda)$. It needs to be supplemented with the initial condition $\chi(t,\Lambda_0)=1$ for all times $t$. The tRG flow equations read
\begin{widetext}
%\begin{flalign}
%\frac{\partial P_\ph(t,\Lambda)}{\partial \Lambda} &= \chi(t,\Lambda) \left\{  2 \int_\s d^d \vec{p} ~p_x  \alpha_{\vec{p}} \frac{ \partial \alpha_{\vec{p}} }{\partial \Lambda} 
%- \int_\f d^{d-1}\vec{k} ~ k_x \Bigl[ |\lambda_{\vec{k}}(t)|^2 + 2 \alpha_{\vec{k}} {\rm Re} \lambda_{\vec{k}}(t) \Bigr] \right\},  &  ~ P_\ph(t,\Lambda_0) = P_\ph^\MF, \qquad \label{eq:tRGflowPphLbdat} \\
% \frac{\partial \chi(t,\Lambda) }{\partial \Lambda} &= 2 \mathcal{M}_{x x}^{-1} \chi(t,\Lambda)  \int_\f d^{d-1} \vec{k} ~ k_x^2 \frac{\alpha_{\vec{k}}}{\Omega_{\vec{k}}} \l {\rm Re} \lambda_{\vec{k}}(t) + \alpha_{\vec{k}} \r, &  \chi(t,\Lambda_0)=1. \qquad \label{eq:tRGflowChiLbdat}
%\end{flalign}
\begin{equation}
\frac{\partial P_\ph(t,\Lambda)}{\partial \Lambda} = \chi(t,\Lambda) \left\{  2 \int_\s d^d \vec{p} ~p_x  \alpha_{\vec{p}} \frac{ \partial \alpha_{\vec{p}} }{\partial \Lambda} 
- \int_\f d^{d-1}\vec{k} ~ k_x \Bigl[ |\lambda_{\vec{k}}(t)|^2 + 2 \alpha_{\vec{k}} {\rm Re} \lambda_{\vec{k}}(t) \Bigr] \right\},  \qquad  ~ P_\ph(t,\Lambda_0) = P_\ph^\MF,  \label{eq:tRGflowPphLbdat}
\end{equation}
\begin{equation}
 \frac{\partial \chi(t,\Lambda) }{\partial \Lambda} = 2 \mathcal{M}_{x x}^{-1} \chi(t,\Lambda)  \int_\f d^{d-1} \vec{k} ~ k_x^2 \frac{\alpha_{\vec{k}}}{\Omega_{\vec{k}}} \l {\rm Re} \lambda_{\vec{k}}(t) + \alpha_{\vec{k}} \r, \qquad  \chi(t,\Lambda_0)=1.  \label{eq:tRGflowChiLbdat}
\end{equation}
\end{widetext}
A similar set of equations can be derived for the total phonon number in the polaron cloud, see Sec.~\ref{eq:tRGderivationsFlowEq}. In Appendix \ref{apdx:fullTimeDeptRG} we generalize these tRG flow equations to deal with explicitly time-dependent Hamiltonians.

%%%%%%%%%%%%%%%%%%%%%%%%%%%%%%%%%%%%%%%%%%%%%%%%%%%%%
\section{Results}
\label{sec:Results}
%%%%%%%%%%%%%%%%%%%%%%%%%%%%%%%%%%%%%%%%%%%%%%%%%%%%%

Now we present results for polaron dynamics, relevant to recent experiments with ultracold atoms \cite{Spethmann2012,Hohmann2015,Jorgensen2016PRL,Hu2016PRL}. In Subsection \ref{subsec:ResPolaronFormation} we calculate polaron trajectories after a sudden interaction quench, as described in \ref{subsec:DynamicsOfPolaronFormation}. For light impurities, $M \leq m_{\rm B}$, and strong interactions, $\alpha \geq 1$, our results deviate substantially from the time-dependent MF predictions in Ref.~\cite{Shashi2014RF}. At the same time, effects from two-phonon terms are captured almost entirely by an effective Fr\"ohlich Hamiltonian when the renormalized coupling from Eq.~\eqref{eqaIBFroehlichEff} is used. 

In Subsection \ref{subsec:ResSpectralFunction}, we calculate the spectral function of the impurity. For strong interactions, we predict a shift of the spectral weight to higher energies, accompanied by the development of a gap-like structure with strongly suppressed spectral weight at low energies above the polaron peak. This effect is reminiscent of the dark continuum predicted in strongly interacting Fermi polarons \cite{Goulko2016}. It is also much more pronounced than expected from time-dependent MF calculations \cite{Shashi2014RF}.

%%%%%%%%%%%%%%%%%%%%%%%%%%%%%%%%%%%%%%%%%%%%%%%%%%%%%
\begin{figure}[b!]
\centering
\epsfig{file=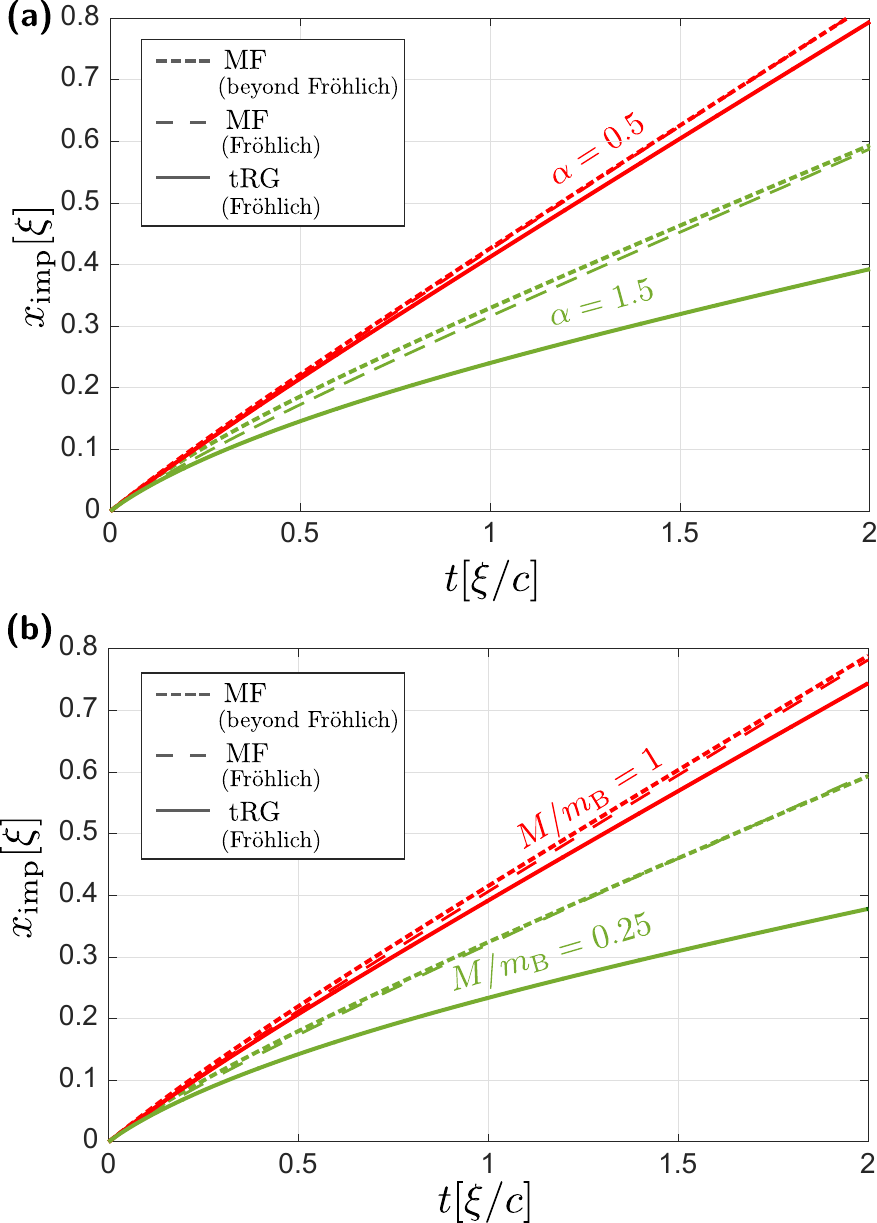, width=0.46\textwidth} $\quad$
\caption{Polaron trajectories are shown after an interaction quench at time $t=0$, from non-interacting to different values of $\alpha$. The initial impurity velocity was $P/M=0.5 c$, and we used a sharp UV cut-off at $\Lambda_0=20/\xi$ in the calculations. In (a) we compare the result for different values of the final coupling $\alpha$, at a mass ratio of $M/m_{\rm B}=0.5$. In (b) we set $\alpha=1$ for all curves and varied the mass ratio $M/m_{\rm B}$. We compare tRG simulations and MF results for the Fr\"ohlich model with MF calculations including beyond-Fr\"ohlich effects.}
\label{fig:PolaronTrajectory3}
\end{figure}
%%%%%%%%%%%%%%%%%%%%%%%%%%%%%%%%%%%%%%%%%%%%%%%%%%%%%

%%%%%%%%%%%%%%%%%%%%%%%%%%%%%%%%%%%%%%%%%%%%%%%%%%%%%
\begin{figure}[t!]
\centering
\epsfig{file=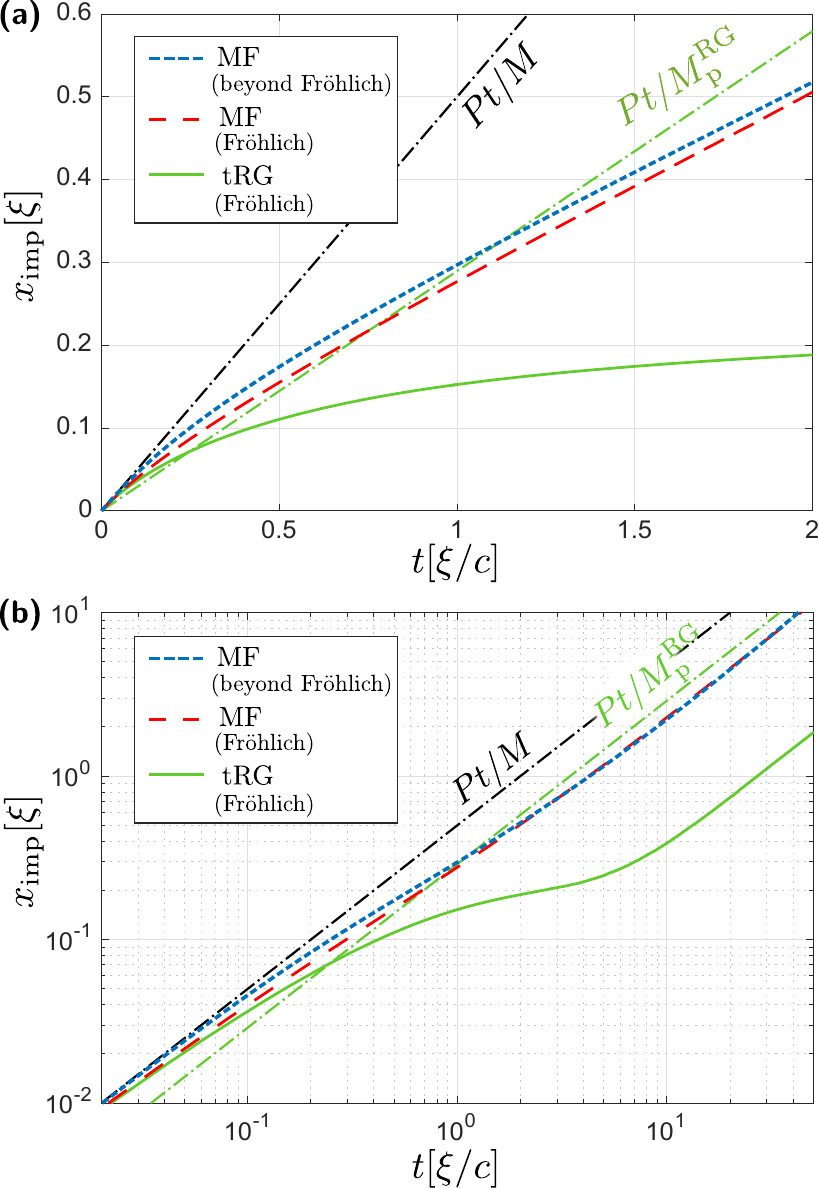, width=0.48\textwidth} $\qquad$
\caption{The polaron trajectory is shown after an interaction quench from $\alpha=0$ to $\alpha=2$ at time $t=0$. The initial impurity velocity was $P/M=0.5 c$, and the mass ratio is $M/m_{\rm B}=0.5$. We used a sharp UV cut-off at $\Lambda_0=20/\xi$ in the calculations. In (b) the same data as in (a) is shown, but on a double-logarithmic scale.}
\label{fig:PolaronTrajectory2}
\end{figure}
%%%%%%%%%%%%%%%%%%%%%%%%%%%%%%%%%%%%%%%%%%%%%%%%%%%%%

%%%%%%%%%%%%%%%%%%%%%%%%%%%%%%%%%%%%%%%%%%%%%%%%%%%%%
\begin{figure*}[t!]
\centering
\epsfig{file=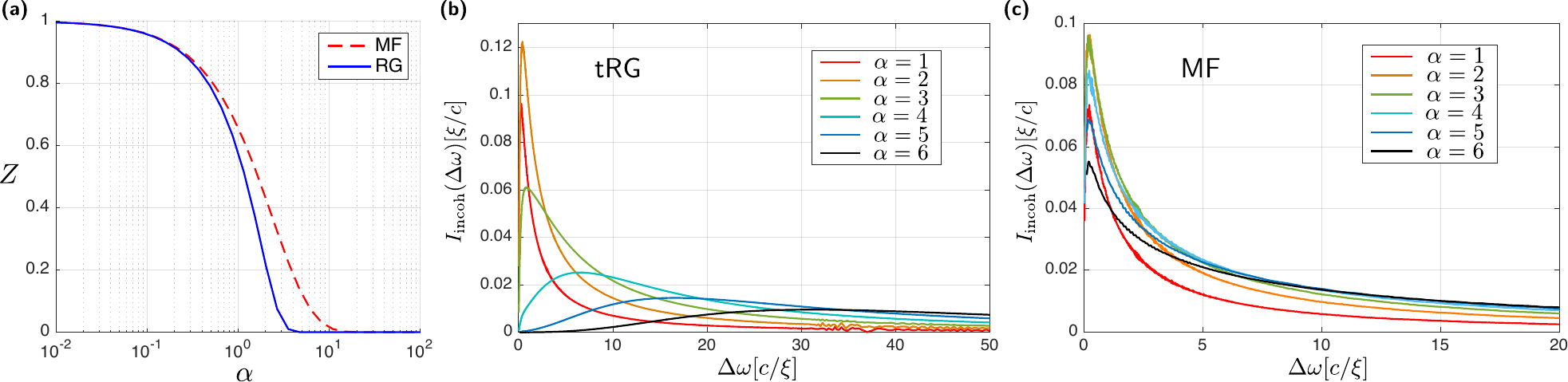, width=\textwidth}
\caption{(a) The polaron quasiparticle weight $Z$ is calculated as a function of the coupling strength $\alpha$ in the Fr\"ohlich Hamiltonian. In (b) and (c) the incoherent part of the spectral function is shown, calculated from tRG (a) and MF (b) respectively. Energies are measured as a difference $\Delta \omega$ from the polaron ground state energy $E_0$ where the coherent delta-peak $I_{\rm coh}(\omega)=Z \delta(\omega-E_0)$ is located. Note the different scales in (b) and (c). In all curves we have chosen $M/m_{\rm B}=0.26$.}
\label{fig:RFspectra1}
\end{figure*}
%%%%%%%%%%%%%%%%%%%%%%%%%%%%%%%%%%%%%%%%%%%%%%%%%%%%%

%%%%%%%%%%%%%%%%%%%%%%%%%%%%%%%%%%%%%%%%%%%%%%%%%%%%%
\begin{figure*}[t!]
\centering
\epsfig{file=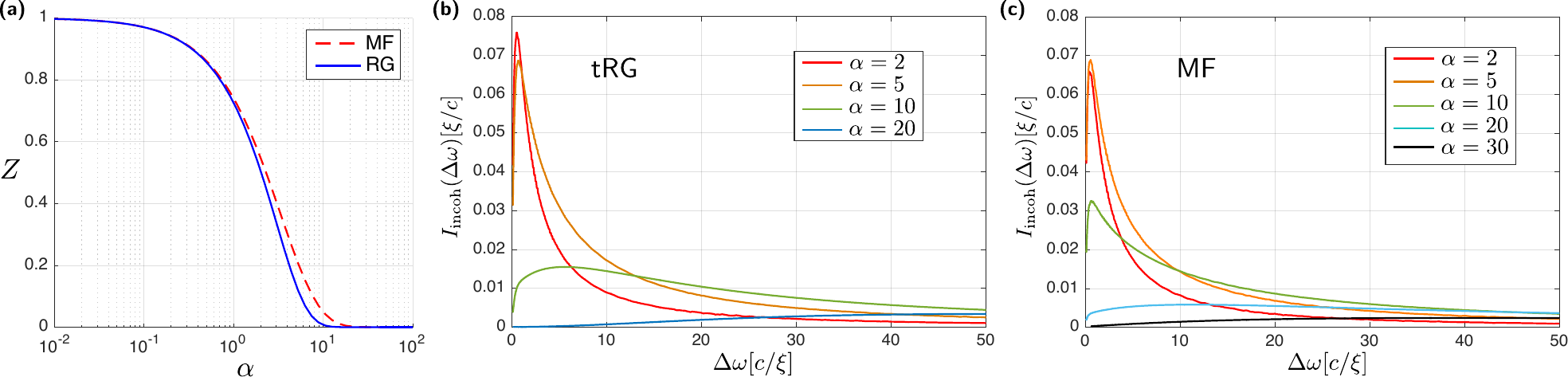, width=\textwidth}
\caption{The same data as in Fig.~\ref{fig:RFspectra1} is shown, but calculated for a larger mass ratio of $M/m_{\rm B}=1$. (b) corresponds to tRG results and (c) to MF theory. They resemble each other much more closely in this case of a heavier impurity than in Fig.~\ref{fig:RFspectra1}.}
\label{fig:RFspectra2}
\end{figure*}
%%%%%%%%%%%%%%%%%%%%%%%%%%%%%%%%%%%%%%%%%%%%%%%%%%%%%

%%%%%%%%%%%%%%%%%%%%%%
\subsection{Dynamics of polaron formation}
\label{subsec:ResPolaronFormation}
%%%%%%%%%%%%%%%%%%%%%%
We start by comparing polaron trajectories for different values of the coupling constant $\alpha$ after the quench in Fig.~\ref{fig:PolaronTrajectory3} (a). For weak couplings, $\alpha=0.5$, we find that the tRG approach follows the time-dependent MF result and the corrections due to quantum fluctuations are small. Notably, solving tRG flow equations is numerically less demanding than solving the coupled MF equations. Therefore the good agreement at weak couplings not only serves as a benchmark of our method but also enables us to solve polaron dynamics more efficiently in the weak coupling regime. 

Our MF and tRG calculations are performed for the Fr\"ohlich Hamiltonian from Eq.~\eqref{eq:HFroh}. We compare them to results of the time-dependent MF approach \cite{Shashi2014RF,Shchadilova2016PRL} applied to the beyond-Fr\"ohlich Hamiltonian, Eq.~\eqref{eq:HbF}. In this case, the microscopic scattering length $a_{\rm IB}$ is chosen such that the same effective coupling strength $\alpha = \alpha^*(a_{\rm IB})$ is obtained which is used in the Fr\"ohlich Hamiltonian, see Eq.~\eqref{eqaIBFroehlichEff}. 

We find from Fig.~\ref{fig:PolaronTrajectory3} (a) that corrections of the tRG to the time-dependent MF results start to become sizable around $\alpha \approx 1$. We observe a quick deceleration of the impurity, which can be intuitively understood by noting that the effective mass of strongly coupled polarons in equilibrium is enhanced by quantum fluctuations. For a wide range of couplings $\alpha$, we find that beyond-Fr\"ohlich effects are very well captured by the use of the renormalized coupling constant in the Fr\"ohlich Hamiltonian. 

In Fig.~\ref{fig:PolaronTrajectory3} (b) we use $\alpha=1$ and compare the resulting impurity trajectories for different values of the impurity-to-boson mass ratio $M/m_{\rm B}$. As expected, in the limit $M \gg m_{\rm B}$ where the time-dependent MF theory becomes exact, the predictions of both approaches coincide. For light impurities, on the other hand, quantum fluctuations lead to strong corrections to the impurity trajectories. We believe that this reflects the large mass renormalization of the polaron ground states in this regime \cite{Grusdt2015RG,Grusdt2016RG}. 

In Fig.~\ref{fig:PolaronTrajectory2} we show a polaron trajectory calculated for an even larger final interaction strength of $\alpha=2$ after the quench. The impurity slows down dramatically and deviates from the MF trajectory at short times. At much longer times, the impurity reaches a steady state with a constant velocity, see Fig.~\ref{fig:PolaronTrajectory2} (b). Notably, the final impurity velocity is much smaller than in the case where the interactions are switched on adiabatically. In the latter case, the impurity would slowly turn into a polaron of mass $M_{\rm p}^{\rm RG}$, with a velocity $P/M_{\rm p}^{\rm RG}$. The small polaron velocity after the quench thus requires emission of many phonons, which carry away part of the initial impurity momentum $\vec{P}$. A similar behavior is predicted by time-dependent MF theory, although the effect is much less pronounced in this case, see Fig.~\ref{fig:PolaronTrajectory2} and Ref.~\cite{Shashi2014RF}. 

From the long-time dynamics shown in Fig.~\ref{fig:PolaronTrajectory2} (b) we note that two-phonon terms do not change the final velocity of the impurity substantially. At short times their effect is more pronounced, leading to a faster polaron than expected from the pure Fr\"ohlich model. This observation demonstrates that the mapping introduced in Eq.~\eqref{eqaIBFroehlichEff} allows explaination of the long-time dynamics of strongly coupled Bose polarons on the attractive side of a Feshbach resonance with an effective Fr\"ohlich model. 

%%%%%%%%%%%%%%%%%%%%%%%%%%%%%%%%%%%%%%%%%%%%%%%%%%%%%
\begin{figure*}[t!]
\centering
\epsfig{file=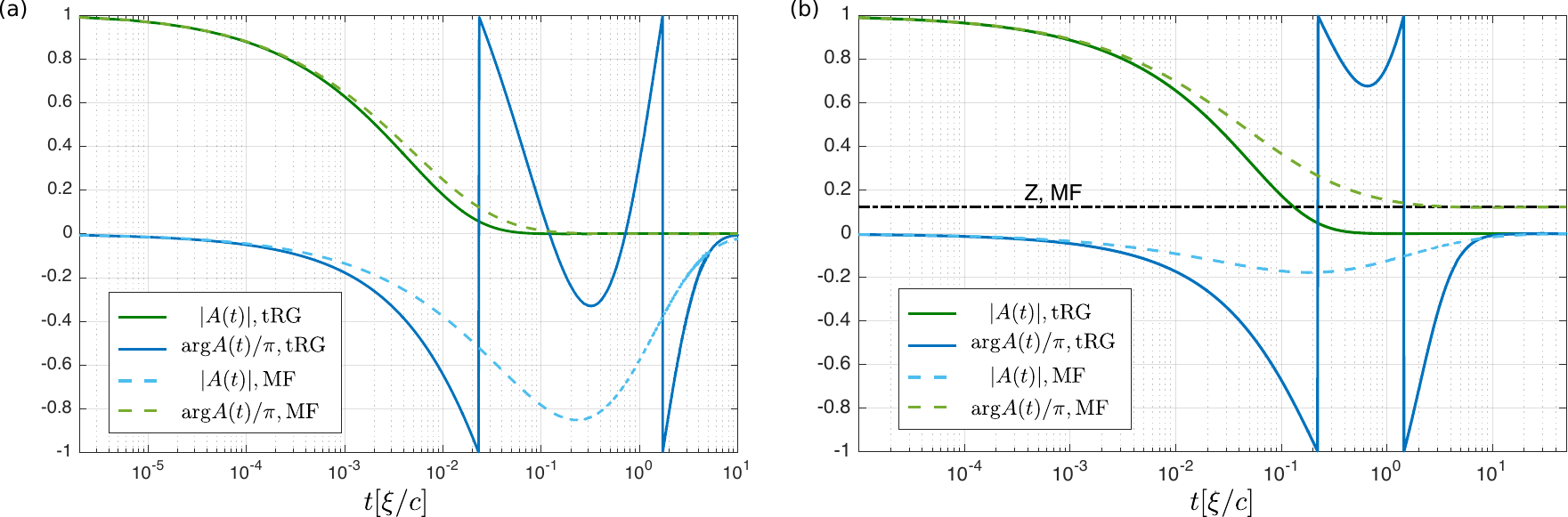, width=0.98\textwidth}
\caption{The time-dependent overlap $A(t)$, defined in Eq.~\eqref{eq:defAt}, is shown for $M/m_{\rm B}=0.26$ at $\alpha=5$ in (a) and for $M/m_{\rm B}=1$ at $\alpha=30$ in (b). For long times the complex phase ${\rm arg} A(t)$ approaches an asymptotic form $E_0 t$, where $E_0$ is the polaron ground state energy. This part was subtracted and we only show how the asymptotic behavior is approached.}
\label{fig:timDepOverlap}
\end{figure*}
%%%%%%%%%%%%%%%%%%%%%%%%%%%%%%%%%%%%%%%%%%%%%%%%%%%%%

Finally, in Fig.~\ref{fig:Intro} (c), we calculate the polaron trajectory in the strong coupling regime, $\alpha=2.1$. In this case, we observe a non-monotonic behavior at intermediate times, where the impurity wavers back. At long times we find similar behavior as presented in Fig.~\ref{fig:PolaronTrajectory2} (b). This wavering demonstrates that quantum fluctuations not only modify the impurity trajectories on a quantitative level, but they also introduce qualitative changes in comparison to MF results. In the regime under consideration, the coupling strength $\alpha$ of the effective Fr\"ohlich Hamiltonian is still sufficiently weak that the effects of two-phonon terms are almost entirely captured by the introduction of the renormalized coupling constant according to Eq.~\eqref{eqaIBFroehlichEff}. Thus we expect that non-monotonic polaron trajectories are not an artifact of the effective Fr\"ohlich model, but can be observed experimentally for strongly coupled Bose polarons near a Feshbach resonance. The discrepancy between MF and tRG is so striking that we expect quantum fluctuations to play an important role in this strongly interacting far-from-equilibrium regime. 

For even larger couplings and sufficiently light impurities, we find that effects of quantum fluctuations are further enhanced. Because it is unclear how reliable this approach is for very strong couplings, a detailed study of the regime $\alpha \gg 1$ will be done in future work.

%%%%%%%%%%%%%%%%%%%%%%%%%%%%%%%%%%%%%%%%%%%%%%%%%%%%%
\begin{figure}[b!]
\centering
\epsfig{file=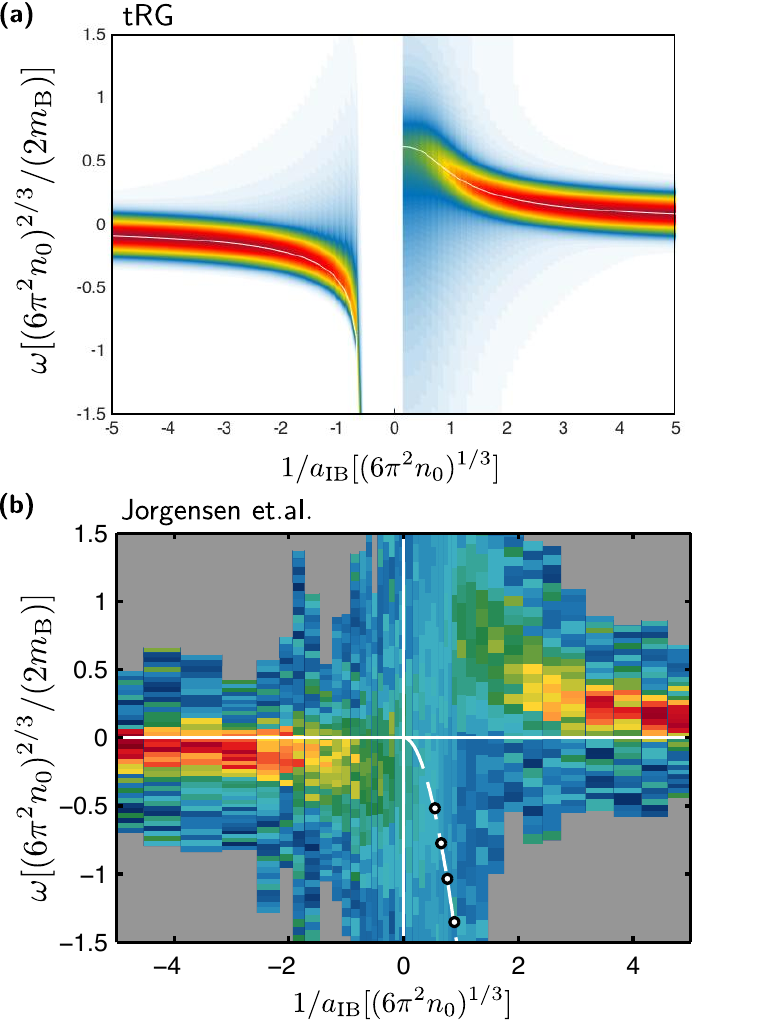, width=0.48\textwidth}
\caption{(a) The spectral function is calculated after including two-phonon terms beyond the Fr\"ohlich Hamiltonian, but still using the Bogoliubov approximation of non-interacting phonons (see Ref.~\cite{Grusdt2017RG}). We used the same parameters as in the experimental observation of Bose polarons \cite{Jorgensen2016PRL}, $M/m_{\rm B}=1$ and $n_0=2.3 \times10^{14} {\rm cm}^{-3}$. A UV cut-off $\Lambda_0=1/60 a_0$ corresponding to the inverse effective range estimated in Ref.~\cite{Jorgensen2016PRL} was used ($a_0$ is the Bohr radius).  We performed calculations at vanishing polaron momentum, $P=0$, and included the same Fourier broadening as discussed in Ref.~\cite{Jorgensen2016PRL}. For comparison, the measured spectrum is shown in (b), taken from Ref.~\cite{Jorgensen2016PRL}.}
\label{fig:compJorgensen}
\end{figure}
%%%%%%%%%%%%%%%%%%%%%%%%%%%%%%%%%%%%%%%%%%%%%%%%%%%%%

%%%%%%%%%%%%%%%%%%%%%%
\subsection{Spectral function}
\label{subsec:ResSpectralFunction}
%%%%%%%%%%%%%%%%%%%%%%
Our results for the incoherent part of the polaron spectral function $I(\omega)$ are shown in Fig.~\ref{fig:RFspectra1}. We chose a mass ratio $M/m_{\rm B}=0.26$ (corresponding to a Li-Na mixture \cite{Tempere2009,Scelle2013}) and varied the coupling strength $\alpha$. From the quasiparticle residue $Z$ of the polaron ground state, shown in Fig.~\ref{fig:RFspectra1} (a), we expect a pronounced cross-over from weak to strong coupling around $\alpha \approx 3$, see Refs.~\cite{Tempere2009,Grusdt2015RG,Shchadilova2016}. 

Indeed, below $\alpha \approx 3$ the polaron spectra predicted by tRG, shown in (b), and by MF theory, shown in (c), are very similar. For larger coupling strengths $\alpha \gtrsim 3$ the tRG predicts a substantial shift of spectral weight to higher energies. This can be seen most prominently by analyzing the width $\Delta \omega$ of the incoherent part of the spectrum. While $\Delta \omega \approx c /\xi$ is approximately constant for weak couplings, it increases quickly in the strong coupling regime due to quantum fluctuations. 

In addition, we observe a strong suppression of spectral weight at low energies above the ground state for $\alpha \gtrsim 3$. This effect is reminiscent of the dark continuum predicted in strongly interacting Fermi polarons \cite{Goulko2016}. It is caused by quantum fluctuations and can only be described by tRG as it is absent in MF calculations. The formation of a gap-like structure can be understood by the build-up of correlations between phonons at low energies due to phonon-phonon interactions induced by the mobile impurity. As a result, the effect is completely absent for an infinite-mass impurity \cite{Shashi2014RF}.

In Fig.~\ref{fig:RFspectra2} we present calculations for a heavier impurity with a mass ratio $M/m_{\rm B} = 1$. Here the cross-over from weak to strong coupling is less pronounced. For small couplings ($\alpha \lesssim 5$), MF theory and tRG agree very well. For stronger couplings, both methods predict a shift of spectral weight to higher energies, although for MF theory larger values of $\alpha$ are required to observe this effect. 

In Fig.~\ref{fig:timDepOverlap} we show results in the time domain. Note that the overlaps $A(t)$ can be directly measured using Ramsey interferometry \cite{Knap2012,Cetina2016}. We find that the amplitude $|A(t)|$ decays quickly on a time scale which is slightly faster than $c/\xi$. In the long time limit, $|A(t)| \to Z$ approaches a constant value given by the polaron quasiparticle weight. When quantum fluctuations are included on top of the MF solution, we find that the phase ${\rm arg} A(t)$ starts to oscillate before it approaches its asymptotic form. This does not lead to a pronounced peak in the spectral function as $|A(t)|$ is strongly suppressed on timescales when these oscillations become relevant.

In Fig.~\ref{fig:compJorgensen} we calculate the spectral function for Bose polarons at strong coupling. Parameters relevant to the experiments of Ref.~\cite{Jorgensen2016PRL} are chosen. Here we generalized the tRG flow equations following Ref.~\cite{Grusdt2017RG} and included two-phonon terms beyond the Fr\"ohlich Hamiltonian, which is not difficult for vanishing polaron momentum $\vec{P}=0$. Comparison with the experimental data in Fig.~\ref{fig:compJorgensen} (b), taken directly from Ref.~\cite{Jorgensen2016PRL}, yields very good agreement. While Fourier broadening was included in our calculations, we did not account for trap averaging which is expected to contribute to the observed deviations. 

In the region around the Feshbach resonance we have no theoretical data because the RG becomes unstable due to phonon-phonon interactions, see discussion in Ref.~\cite{Grusdt2017RG}. In this regime, MF theory predicts a broad spectrum due to multiparticle bound states \cite{Shchadilova2016PRL}, and several works concluded that interaction effects in the background Bose gas play an important role \cite{Ardila2015,Grusdt2017RG,Grusdt2016RG1D}. These effects are consistent with the featureless spectrum observed in the experiment. We also note that the tRG approach does not include the molecular bound states on the repulsive side of the Feshbach resonance which have been predicted by time-dependent MF calculations of the Bose polaron problem \cite{Shchadilova2016PRL}, or even more intricate multi-particle Efimov states \cite{Levinsen2015,Sun2017,Sun2017a}.

%%%%%%%%%%%%%%%%%%%%%%%%%%%%%%%%%%%%%%%%%%%%%%%%%%%%%
\section{Dynamical RG approach}
\label{sec:tRG}
%%%%%%%%%%%%%%%%%%%%%%%%%%%%%%%%%%%%%%%%%%%%%%%%%%%%%

In this section we derive the tRG flow equations. To this end, we generalize the RG approach introduced for equilibrium problems in Refs. \cite{Grusdt2015RG,Grusdt2015Varenna}. The following calculations include the extensions suggested in Ref. \cite{Grusdt2016RG} to obtain an RG approach for the Fr\"ohlich model valid for arbitrary coupling strengths. The notations in this section are the same as in Sec.~\ref{sec:Overview}. 

This section contains two parts. In the first part, we formulate the tRG for time-dependent physical observables. In the second part, we generalize the approach to the calculation of time-dependent overlaps required for the spectral function. Using the results from Ref.~\cite{Grusdt2017RG} these calculations can be generalized to include two-phonon terms \cite{Rath2013} and calculate the Bose polaron spectra shown in Figs.~\ref{fig:Intro} (b) and \ref{fig:compJorgensen} (a).

%%%%%%%%%%%%%%%%%%%%%%
\subsection{Physical observables}
\label{sec:PhysicalObservalestRG}
%%%%%%%%%%%%%%%%%%%%%%
Here we calculate the time-dependence of the phonon momentum $\vec{P}_\ph(t)$ and the phonon number $N_\ph(t)$ in the polaron cloud. We consider the physical situation described in Sec.~\ref{subsec:DynamicsOfPolaronFormation} where the initial state corresponds to the phonon vacuum, $\ket{\psi_0}=\ket{0}$, and impurity-boson interactions are switched on suddenly at time $t=0$. 

Both $\hat{N}_\ph = \int d^d \vec{k} ~ \ad_{\vec{k}} \a_{\vec{k}}$ and $\hat{\vec{P}}_\ph = \int d^d \vec{k} ~ \vec{k} \ad_{\vec{k}} \a_{\vec{k}}$ commute with the LLP transformation \eqref{eq:LLPdef}. For applying the MF shift in Eq. \eqref{eq:UMF} we use that
\begin{equation}
\hat{U}_{\rm MF}^\dagger \ad_{\vec{k}} \a_{\vec{k}} \hat{U}_{\rm MF} = \G_{\vec{k}}(\Lambda_0) + \l \alpha_{\vec{k}}^\MF \r^2.
\end{equation}
The initial state in the frame of quantum fluctuations around the MF polaron reads $\ket{\tilde{\psi}_0} = \hat{U}_\MF^\dagger \ket{0} = \ket{- \alpha_{\vec{\kappa}}^\MF}$. Here $\ket{- \alpha_{\vec{\kappa}}^\MF}$ is a short-hand notation for $\prod_{\vec{\kappa}} \ket{- \alpha_{\vec{\kappa}}^\MF}$. We find the following expressions in the basis of quantum fluctuations, $N_\ph(t) = N_\ph^\MF + \Delta N_\ph(t)$ and $\vec{P}_\ph(t) = \vec{P}_\ph^\MF + \Delta \vec{P}_\ph(t)$, where
\begin{flalign}
 \Delta N_\ph(t) &= \int^{\Lambda_0} d^d \vec{k} ~ \bra{-\alpha_{\vec{\kappa}}^\MF} e^{i \tilde{\mathcal{H}} t } \G_{\vec{k}}(\Lambda_0) e^{- i \tilde{\mathcal{H}} t } \ket{- \alpha_{\vec{\kappa}}^\MF}, \label{eq:NphtUniversal} \\
  \Delta \vec{P}_\ph(t) &= \int^{\Lambda_0} d^d \vec{k} ~ \vec{k} \bra{-\alpha_{\vec{\kappa}}^\MF} e^{i \tilde{\mathcal{H}} t } \G_{\vec{k}}(\Lambda_0) e^{-i \tilde{\mathcal{H}} t } \ket{- \alpha_{\vec{\kappa}}^\MF}. \label{eq:PphtUniversal}
\end{flalign}
Now we will derive the tRG flow for the observable $\hat{o}_{\vec{k}} = \G_{\vec{k}}$, cf. Eq.~\eqref{eq:OtFlucFrame}, from which expression for $\Delta N_\ph(t)$ and $ \Delta \vec{P}_\ph(t)$ can easily be derived.

%%%%%%%%%%%
\subsubsection{tRG step}
\label{sec:tRGstep}
%%%%%%%%%%%
The RG transformation $\hat{U}_\Lambda$ discussed in Sec.~\ref{subsec:tRGOverview} consists of two parts in the extended RG scheme of Ref. \cite{Grusdt2016RG}, $\hat{U}_\Lambda = \hat{W}_\Lambda \hat{V}_\MF(\Lambda) $. We will discuss their effects one after the other now.

We start every tRG step by introducing the unitary transformation $\hat{W}_\Lambda$ derived for the equilibrium problem in the perturbative RG approach \cite{Grusdt2015RG,Grusdt2016RG}. We utilize it to diagonalize fast phonon degrees of freedom in the universal Hamiltonian \eqref{eq:HquantFluc2} at a given UV cut-off $\Lambda$. This transformation is defined by
\begin{equation}
\hat{W}_\Lambda =  \exp \l \int_\f d^d \vec{k} ~ \left[ \F_{\vec{k}}^\dagger \a_{\vec{k}} - \F_{\vec{k}} \ad_{\vec{k}} \right] \r
\label{eq:WMF-RG}
\end{equation}
and describes the displacement of fast phonons by an amount which depends on slow phonon variables, 
\begin{equation}
\F_{\vec{k}} =  \frac{ \alpha_{\vec{k}} }{\Omega_{\vec{k}}}  k_\mu \mathcal{M}_{\mu \nu}^{-1} \int_\s d^d \vec{p} ~  p_\nu  \G_{\vec{p}}  + \mathcal{O}(\Omega_{\vec{k}}^{-2}).
\label{eq:defFk}
\end{equation}
We dropped the arguments in $\alpha_{\vec{k}} = \alpha_{\vec{k}}(\Lambda)$, $\Omega_{\vec{k}} = \Omega_{\vec{k}}(\Lambda)$, $\mathcal{M}_{\mu \nu}=\mathcal{M}_{\mu \nu}(\Lambda)$ and $\G_{\vec{p}} =\G_{\vec{p}}(\Lambda)$ to retain clarity in our notation. 

Now the contribution to the phonon number reads
\begin{multline}
\Delta N_\ph(t) = \int^\Lambda d^d \vec{k} ~  \bra{-\alpha_{\vec{\kappa}}} \hat{W}_\Lambda e^{ i \hat{W}_\Lambda^\dagger \tilde{\mathcal{H}} \hat{W}_\Lambda t} \\ \times  \hat{W}_\Lambda^\dagger \G_{\vec{k}}  \hat{W}_\Lambda e^{- i \hat{W}_\Lambda^\dagger \tilde{\mathcal{H}} \hat{W}_\Lambda t} \hat{W}_\Lambda^\dagger \ket{-\alpha_{\vec{\kappa}}}.
\label{eq:OtDerivation1}
\end{multline}
Here we assumed that the initial state is given by the product of coherent states $\ket{-\alpha_{\vec{\kappa}}}$, which is true initially because $\alpha_{\vec{k}}(\Lambda_0)=\alpha_{\vec{k}}^\MF$. We will show below that the amplitude $\alpha_{\vec{k}}$ of the initial state is exactly the renormalized MF-type amplitude flowing in the extended tRG scheme, i.e. $\alpha_{\vec{k}}=-V_k/\Omega_{\vec{k}}$.

Next we separate the Hamiltonian into fast and slow phonon contributions $\Ht_\s$ and $\Ht_\f$, respectively, as well as couplings between them $\Ht_\sf$. As in the equilibrium RG \cite{Grusdt2015RG,Grusdt2016RG}, we find that $\hat{W}_\Lambda$ diagonalizes fast phonons,
\begin{equation}
 \hat{W}_\Lambda^\dagger \tilde{\cal H} \hat{W}_\Lambda = \int_\f d^d \vec{k} ~ \l \Omega_{\vec{k}} + \hat{\Omega}_\s(\vec{k}) \r \ad_{\vec{k}} \a_{\vec{k}} + \Ht_\s + \delta \Ht_\s.
 \label{eq:renHamNonEqMAIN}
\end{equation}
The slow phonon Hamiltonian is renormalized by
\begin{multline}
\delta \Ht_\s =  - \int_\f d^d \vec{k} ~ \frac{1}{\Omega_{\vec{k}}} \left[ \alpha_{\vec{k}}   k_\mu \mathcal{M}_{\mu \nu}^{-1} \int_s d^d \vec{p} ~ p_\nu \G_{\vec{p}} \right]^2 \\
+ \int_\f d^d \vec{k} ~ \frac{k_\mu  \mathcal{M}_{\mu \nu}^{-1} k_\nu }{2} \alpha_{\vec{k}}^2  + \mathcal{O}(\Omega_{\vec{k}})^{-2},
\label{eq:renHs}
\end{multline}
as in the equilibrium RG. The frequency renormalization of fast phonons by 
\begin{equation}
\hat{\Omega}_\s(\vec{k}) = k_\mu \mathcal{M}_{\mu \nu}^{-1} \int_\s d^d \vec{p} ~ p_\nu \G_{\vec{p}}
\label{eq:defOmegaSk}
\end{equation}
leads to additional renormalization of the slow-phonon Hamiltonian. This term is specific to the tRG. 

We treat the new terms $\hat{\Omega}_\s(\vec{k})$ perturbatively. From a Trotter decomposition of the time-evolution we obtain the leading-order contribution
\begin{multline}
e^{- i \Ht t} =  e^{-i \int_\f d^d \vec{k} ~ \Omega_{\vec{k}} \ad_{\vec{k}} \a_{\vec{k}}} \Biggl[ e^{ - i \l \Ht_\s + \delta \Ht_\s \r t} - i \int_0^t d \tau  \\ \times e^{ - i \l \Ht_\s + \delta \Ht_\s \r \l t - \tau \r}  \int_\f d^d \vec{k} ~ \ad_{\vec{k}} \a_{\vec{k}} \hat{\Omega}_\s(\vec{k}) e^{ -  i \l \Ht_\s + \delta \Ht_\s \r \tau } \Biggr].
\label{eq:trotterPert}
\end{multline}

In Eq.~\eqref{eq:OtDerivation1} we need to evaluate the last expression in the state $\hat{W}_\Lambda^\dagger \ket{-\alpha_{\vec{\kappa}}}$. To this end we note that the slow-phonon coherent states $\ket{- \alpha_{\vec{p}}}$ are eigenstates of $\G_{\vec{p}}$, 
\begin{equation}
\G_{\vec{p}} \ket{- \alpha_{\vec{p}}}_\s = - \l \alpha_{\vec{p}} \r^2 ~ \ket{- \alpha_{\vec{p}}}_\s,
 \label{eq:GammaEigenstatesMAIN}
\end{equation}
as one easily verifies. Hence $\hat{W}_\Lambda^\dagger \ket{-\alpha_{\vec{k}}}_\f \ket{-\alpha_{\vec{p}}}_\s = \ket{ \lambda_{\vec{k}}}_\f \ket{-\alpha_{\vec{p}}}_\s$ yields a coherent state with the fast-phonon amplitude $\lambda_{\vec{k}} =  f_{\vec{k}} - \alpha_{\vec{k}}$,
\begin{equation}
f_{\vec{k}} = - \frac{\alpha_{\vec{k}}}{\Omega_{\vec{k}}}  k_\mu \mathcal{M}_{\mu \nu}^{-1} \int_\s d^d \vec{p} ~  p_\nu  (\alpha_{\vec{p}})^2 .
\label{eq:def_fkMAIN}
\end{equation}

\paragraph{Slow phonons -- Hamiltonian renormalization}
Now we proceed differently for fast and slow phonon contributions to the phonon number. For slow phonons we use that $\hat{W}_\Lambda^\dagger \G_{\vec{p}}  \hat{W}_\Lambda = \G_{\vec{p}} + \mathcal{O}(\Omega_{\vec{k}}^{-2})$, allowing us to approximate $\hat{W}_\Lambda^\dagger \G_{\vec{p}}  \hat{W}_\Lambda \approx \G_{\vec{p}}$ (we work accurately to order $\Omega_{\vec{k}}^{-1}$). Noting that
\begin{equation}
\bra{\lambda_{\vec{k}}} e^{i \int_\f d^d \vec{k} ~ \ad_{\vec{k}} \a_{\vec{k}} \Omega_{\vec{k}} t} \ad_{\vec{k}} \a_{\vec{k}} e^{- i \int_\f d^d \vec{k} ~ \ad_{\vec{k}} \a_{\vec{k}} \Omega_{\vec{k}} t} \ket{\lambda_{\vec{k}}} = |\lambda_{\vec{k}}|^2
\label{eq:nkExpctLambdakt}
\end{equation}
and using Eq.~\eqref{eq:trotterPert} we arrive at
\begin{multline}
\Delta N_\ph(t) |_\s = \int_\s d^d \vec{p}  ~ ~_\s \bra{-\alpha_{\vec{p}}} e^{ i \l \Ht_\s + \delta \Ht_\s  + \int_\f d^d \vec{k} ~ |\lambda_{\vec{k}}|^2 \hat{\Omega}_\s(\vec{k})  \r t} \\
 \times \G_{\vec{p}} e^{- i \l \Ht_\s + \delta \Ht_\s  + \int_\f d^d \vec{k} ~ |\lambda_{\vec{k}}|^2 \hat{\Omega}_\s(\vec{k})  \r t} \ket{-\alpha_{\vec{p}}}_\s.
\label{eq:OtDerivationS}
\end{multline}
I.e. the slow-phonon Hamiltonian is renormalized to
\begin{multline}
\delta \Ht_\s' =  - \int_\f d^d \vec{k} ~ \frac{1}{\Omega_{\vec{k}}} \left[ \alpha_{\vec{k}}   k_\mu \mathcal{M}_{\mu \nu}^{-1} \int_s d^d \vec{p} ~ p_\nu \G_{\vec{p}} \right]^2 \\
+   \int_\f d^d \vec{k} ~ |\lambda_{\vec{k}}|^2  k_\mu \mathcal{M}_{\mu \nu}^{-1} \int d^d \vec{p} ~ p_\nu \G_{\vec{p}} \\
+ \int_\f d^d \vec{k} ~ \frac{k_\mu  \mathcal{M}_{\mu \nu}^{-1} k_\nu }{2} \alpha_{\vec{k}}^2 + \mathcal{O}(\Omega_{\vec{k}}^{-2}).
\label{eq:renHs2}
\end{multline}

Comparison to the universal Hamiltonian \eqref{eq:HquantFluc2} shows that the term in the first line of \eqref{eq:renHs2} gives rise to mass renormalization. The renormalized expression after the RG step reads
\begin{equation}
\tilde{\mathcal{M}}_{\mu \nu} ^{-1} = \mathcal{M}_{\mu \nu}^{-1} - 2 \mathcal{M}_{\mu \lambda}^{-1}  \int_\f d^d \vec{k} ~ \frac{\alpha_{\vec{k}}^2 }{\Omega_{\vec{k}}} k_\lambda k_\sigma  ~\mathcal{M}_{\sigma \nu}^{-1},
\end{equation}
leading to the RG flow equation \eqref{eq:RGflowMass} for the tensorial mass. The last line in Eq.~\eqref{eq:renHs2} gives rise to a tRG flow of the zero-point energy.

Most interesting to us is the term in the middle line of Eq.~\eqref{eq:renHs2}, which causes a tRG flow of the phonon momentum. To show this, we bring the renormalized slow-phonon Hamiltonian $\Ht_\s' = \Ht_\s + \delta \Ht_\s'$  to the following normal-ordered form,
\begin{multline}
\Ht_\s'  = E_0' (\Lambda) + \int_\s d^d \vec{p} ~ d^d \vec{p}' ~ \frac{1}{2}  p_\mu \tilde{\mathcal{M}}_{\mu \nu}^{-1} p_\nu'  ~ : \G_{\vec{p}} \G_{\vec{p}'}  : \\
 + \int_\s d^d \vec{p} \left[ \ad_{\vec{p}} \a_{\vec{p}} \Omega_{\vec{p}} + W_{\vec{p}} \G_{\vec{p}} \right].
\label{eq:HquantFlucForMFshiftRG}
\end{multline}
Here we find 
\begin{equation}
E_0' (\Lambda) = E_0(\Lambda) + \frac{1}{2}  \int_\s d^d \vec{p} ~ p_\mu \left[  \tilde{\mathcal{M}}_{\mu \nu}^{-1} -  \mathcal{M}_{\mu \nu}^{-1}  \right] p_\nu \l \alpha_{\vec{p}} \r^2,
\end{equation}
which yields the tRG flow equation \eqref{eq:RGflowZeroPtEnergy} for the zero-point energy. In Eq.~\eqref{eq:HquantFlucForMFshiftRG} we introduced
\begin{equation}
W_{\vec{p}} = \frac{1}{2} p_\mu \left[  \tilde{\mathcal{M}}_{\mu \nu}^{-1} -  \mathcal{M}_{\mu \nu}^{-1}  \right] p_\nu + p_\mu \mathcal{M}_{\mu \nu}^{-1} \int_\f d^d \vec{k} ~ |\lambda_{\vec{k}}|^2 k_\nu,
\end{equation}
which is of order $\mathcal{O}(\delta \Lambda)$. Therefore we may restrict ourselves to a perturbative treatment of such terms to first order in $\delta \Lambda$ now.

Next, we consider the slow phonon Hamiltonian \eqref{eq:HquantFlucForMFshiftRG} in the basis of quantum fluctuations around its mean-field saddle point, which was the key novelty in the extended RG scheme of Ref. \cite{Grusdt2016RG}. To this end we apply a unitary mean-field shift,
\begin{equation}
\hat{V}_\MF (\Lambda) = \exp \l \int_\s d^d \vec{p} ~  \delta \alpha_{\vec{p}} ~  \ad_{\vec{p}} -  \hc \r,
\label{eq:VMFdef}
\end{equation}
which leads to a tRG flow of the coherent amplitudes $\alpha_{\vec{k}}$ appearing in the definition of operators $\G_{\vec{k}}$ in the universal Hamiltonian \eqref{eq:HquantFluc2}. In Ref. \cite{Grusdt2016RG} we have shown that picking the saddle point solution $\delta \alpha_{\vec{p}}$ leads to a tRG flow of the renormalized dispersion which is given by $\Omega_{\vec{p}}' = \Omega_{\vec{p}} \l 1 - \delta \alpha_{\vec{p}} / \alpha_{\vec{p}} \r$. Because $\delta \alpha_{\vec{p}} = \mathcal{O}(\delta \Lambda)$, this proofs that the tRG flow of the coherent amplitudes is given by
\begin{equation}
\alpha_{\vec{p}}(\Lambda-\delta \Lambda) := \alpha_{\vec{p}}(\Lambda) + \delta \alpha_{\vec{p}} = - \frac{V_p}{\Omega_{\vec{p}}(\Lambda-\delta \Lambda)}.
\end{equation}

We derive the tRG flow of the the renormalized dispersion relation $\Omega_{\vec{k}}(\Lambda)$ as in Ref.~\cite{Grusdt2016RG} and obtain
\begin{equation}
\Omega_{\vec{p}}' = \Omega_{\vec{p}} + W_{\vec{p}} + p_x \tilde{\mathcal{M}}_{xx}^{-1} \zeta_x,
\label{eq:OmegapExtDynRG}
\end{equation}
where $\zeta_x := 2 \int_\s d^d \vec{p} ~ p_x \alpha_{\vec{p}} \delta \alpha_{\vec{p}}$ describes how much the MF phonon momentum changes due to the tRG flow of the coherent amplitude $\alpha_{\vec{p}}$. As in Ref.~\cite{Grusdt2016RG}, because we work perturbatively in $\delta \alpha_{\vec{p}} = \mathcal{O}(\delta \Lambda)$, it can easily be determined from the MF saddle point equations. We find
\begin{equation}
\zeta_x = - \frac{2 \int_\s d^d \vec{p} ~ p_x \frac{\alpha_{\vec{p}}^2}{\Omega_{\vec{p}}} W_{\vec{p}}}{1 + 2 \mathcal{M}_{xx}^{-1} \int_\s d^d \vec{p} ~ p_x^2 \frac{\alpha_{\vec{p}}^2}{\Omega_{\vec{p}}}}.
\end{equation}
We plug this result into Eq.~\eqref{eq:OmegapExtDynRG} and confirm that the mass $\mathcal{M}_{\mu \nu}$ in the dispersion flows as described by the tRG flow equation \eqref{eq:RGflowMass}. Then it is easy to derive also the tRG flow equation \eqref{eq:MFRGflowkappaX} for $\kappa_x$. 

Finally we return to Eq.~\eqref{eq:OtDerivationS} where we introduce the unitary MF rotation \eqref{eq:VMFdef}. In the new basis we obtain the fully renormalized Hamiltonian $\Ht(\Lambda-\delta \Lambda) = \hat{V}_\MF^\dagger \Ht_\s' \hat{V}_\MF$. The MF rotation acts on the initial states where the coherent amplitudes are renormalized as expected, $\hat{V}_\MF^\dagger \ket{-\alpha_{\vec{p}}(\Lambda)} = \ket{- \alpha_{\vec{p}}(\Lambda - \delta \Lambda)}$. The operator $\G_{\vec{p}}(\Lambda)$ transforms as
\begin{equation}
\hat{V}_\MF^\dagger \G_{\vec{p}}(\Lambda) \hat{V}_\MF = \G_{\vec{p}}(\Lambda-\delta) + 2 \alpha_{\vec{p}}(\Lambda) \delta \alpha_{\vec{p}}.
\end{equation}
Hence we end up with an expression for $\Delta N_\ph(t)|_\s = \delta N_\ph^\s + \Delta N_\ph'(t)$ consisting of two parts. The first term originates from the tRG flow of the coherent amplitude and it contributes to the phonon number,
\begin{equation}
\delta N_\ph^\s = 2 \int_\s d^d \vec{p} ~ \alpha_{\vec{p}} \delta \alpha_{\vec{p}}.
\label{eq:NphSlow}
\end{equation}
The second term $\Delta N_\ph'(t)$ is of the same algebraic form as the initial expression at cut-off $\Lambda$,
\begin{multline}
\Delta N_\ph'(t) = \int^{\Lambda-\delta \Lambda} d^d \vec{k} ~ \bra{-\alpha_{\vec{\kappa}}(\Lambda-\delta \Lambda)} e^{i \tilde{\mathcal{H}} t } \\ \times
\G_{\vec{k}}(\Lambda-\delta \Lambda) e^{- i \tilde{\mathcal{H}}(\Lambda-\delta \Lambda) t } \ket{- \alpha_{\vec{\kappa}}(\Lambda-\delta \Lambda)}, 
\label{eq:NphtUniversal2} 
\end{multline}
cf. Eq.~\eqref{eq:NphtUniversal}. We can apply subsequent tRG steps to this expression and obtain the entire tRG flow.

\paragraph{Fast phonons -- Flow of observables}
For the evaluation of the fast phonon contribution to the phonon number we need to calculate the effect of the RG transformation $\hat{W}_\Lambda$ on $\G_{\vec{k}}$,
\begin{multline}
\hat{W}_\Lambda^\dagger \G_{\vec{k}} \hat{W}_\Lambda = \l \alpha_{\vec{k}} - \hat{F}_{\vec{k}} \r \l  \a_{\vec{k}} + \ad_{\vec{k}} \r  + \ad_{\vec{k}} \a_{\vec{k}} \\
 - 2 \alpha_{\vec{k}} \hat{F}_{\vec{k}} + \mathcal{O}(\Omega_{\vec{k}}^{-2}).
\end{multline}
Here we used that $\hat{F}_{\vec{k}}^\dagger = \hat{F}_{\vec{k}}$ and $\hat{F}_{\vec{k}} = \mathcal{O}(\Omega_{\vec{k}}^{-1})$.

From Eqs.\eqref{eq:OtDerivation1}, \eqref{eq:GammaEigenstatesMAIN} we thus obtain
\begin{multline}
\Delta N_\ph(t)|_\f =  \int_\f d^d \vec{k}  ~ ~_\f \bra{\lambda_{\vec{k}}} ~_\s \bra{-\alpha_{\vec{p}}} e^{ i \Ht' t}  \Biggl[ \l \alpha_{\vec{k}} - \hat{F}_{\vec{k}} \r \\ \times
 \l  \a_{\vec{k}} + \ad_{\vec{k}} \r  + \ad_{\vec{k}} \a_{\vec{k}} - 2 \alpha_{\vec{k}} \hat{F}_{\vec{k}} \Biggr]  e^{ i \Ht' t} \ket{-\alpha_{\vec{p}}}_\s \ket{\lambda_{\vec{k}}}_\f.
\label{eq:OtDerivationF}
\end{multline}
In the transformed Hamiltonian $\Ht' =  \Ht_\s + \delta \Ht_\s  + \int_\f d^d \vec{k} ~\ad_{\vec{k}} \a_{\vec{k}} \l \Omega_{\vec{k}} + \hat{\Omega}_\s(\vec{k}) \r$ the terms leading to renormalization of the slow phonon Hamiltonian, i.e. $\delta \Ht_\s$ and $\hat{\Omega}_\s(\vec{k})$, can be neglected because they yield corrections to $\Delta N_\ph(t)|_\f$ of order $\mathcal{O}(\delta \Lambda^2)$ only. We may thus write
\begin{equation}
  \Ht' \approx \Ht_\s + \int_\f d^d \vec{k} ~ \ad_{\vec{k}} \a_{\vec{k}} \Omega_{\vec{k}}.
\end{equation}

The fast phonon dynamics in Eq.~\eqref{eq:OtDerivationF} can easily we evaluated because $\hat{F}_{\vec{k}}$ contains only slow phonons, see Eq. \eqref{eq:defFk}. To this end we use that $e^{- i \ad_{\vec{k}} \a \Omega_{\vec{k}} t} \ket{\lambda_{\vec{k}}} = \ket{\lambda_{\vec{k}}(t)}$, where $\ket{\lambda_{\vec{k}}}$ is a coherent state and 
\begin{equation}
\lambda_{\vec{k}}(t) = \lambda_{\vec{k}} e^{- i \Omega_{\vec{k}} t}.
\label{eq:lambdakt}
\end{equation}
We find for the contribution from fast phonons,
\begin{multline}
\Delta N_\ph(t)|_\f =  \int_\f d^d \vec{k} \Bigl[  |\lambda_{\vec{k}}|^2 + 2 \alpha_{\vec{k}} {\rm Re} \lambda_{\vec{k}}(t) \Bigr] +\\ 
+  \int_\s d^d \vec{p}  ~ p_\mu \bra{- \alpha_{\vec{p}}} e^{i \Ht_\s t} \G_{\vec{p}} e^{- i \Ht_\s t} \ket{-\alpha_{\vec{p}}}  \\
 \times 2 \mathcal{M}_{\mu \nu}^{-1} \int_\f d^d \vec{k} ~ k_\nu \frac{V_k}{\Omega_{\vec{k}}^2} \l  {\rm Re} \lambda_{\vec{k}}(t) + \alpha_{\vec{k}}  \r .
\label{eq:DeltaNphFast}
\end{multline}
We can bring the expression in the middle in the usual form required to apply the next tRG step,
\begin{multline}
\bra{- \alpha_{\vec{p}}(\Lambda')} e^{i \Ht(\Lambda') t} \G_{\vec{p}}(\Lambda') e^{- i \Ht(\Lambda') t}  \ket{-\alpha_{\vec{p}}(\Lambda')}
 \\  = \bra{- \alpha_{\vec{p}}} e^{i \Ht_\s t} \G_{\vec{p}} e^{- i \Ht_\s t} \ket{-\alpha_{\vec{p}}}   + \mathcal{O}(\delta \Lambda),
\end{multline}
where $\Lambda' = \Lambda - \delta \Lambda$. This requires only modifications of order $\delta \Lambda^2$ in $\Delta N_\ph(t)|_\f $.

%%%%%%%%%%%
\subsubsection{tRG flow equations}
\label{eq:tRGderivationsFlowEq}
%%%%%%%%%%%
By combining equation \eqref{eq:DeltaNphFast} with the contribution from slow phonons, see Eq. \eqref{eq:NphSlow}, we will now derive tRG flow equations for the phonon momentum and the phonon number. To this end we define 
\begin{widetext}
\begin{equation}
\Delta N_\ph(t,\Lambda) : = \int^{\Lambda} d^d \vec{k} ~ \bra{-\alpha_{\vec{\kappa}}(\Lambda)} e^{i \tilde{\mathcal{H}} t } %\\ \times
\G_{\vec{k}}(\Lambda) e^{- i \tilde{\mathcal{H}}(\Lambda) t } \ket{- \alpha_{\vec{\kappa}}(\Lambda)}, 
\end{equation}
and analogously for the phonon momentum (always directed along $\vec{e}_x$)
\begin{equation}
\Delta P_\ph(t,\Lambda) : = \int^{\Lambda} d^d \vec{k} ~ k_x \bra{-\alpha_{\vec{\kappa}}(\Lambda)} e^{i \tilde{\mathcal{H}} t } %\\ \times
\G_{\vec{k}}(\Lambda) e^{- i \tilde{\mathcal{H}}(\Lambda) t } \ket{- \alpha_{\vec{\kappa}}(\Lambda)}. 
\label{eq:defDeltaPpht}
\end{equation}
From the calculations above we find the following set of coupled flow equations for $\delta N_\ph(t,\Lambda) =  \Delta N_\ph(t,\Lambda-\delta \Lambda) - \Delta N_\ph(t,\Lambda)$ and $\delta P_\ph(t,\Lambda) =  \Delta P_\ph(t,\Lambda-\delta \Lambda) - \Delta P_\ph(t,\Lambda)$,
\begin{flalign}
\delta N_\ph(t,\Lambda) &= - 2 \int_\s d^d \vec{p} ~ \alpha_{\vec{p}} \delta \alpha_{\vec{p}} - \int_\f d^d \vec{k} ~ \Bigl[ |\lambda_{\vec{k}}|^2 + 2 \alpha_{\vec{k}} {\rm Re} \lambda_{\vec{k}}(t)  - \Delta P_\ph(t,\Lambda-\delta \Lambda) 2 \mathcal{M}_{xx}^{-1} k_x \frac{\alpha_{\vec{k}}}{\Omega_{\vec{k}}} \l {\rm Re} \lambda_{\vec{k}}(t) + \alpha_{\vec{k}} \r \Bigr], \label{eq:tRGfullNph}
\end{flalign}
\begin{flalign}
\delta P_\ph(t,\Lambda) &=- 2 \int_\s d^d \vec{p} ~ p_x \alpha_{\vec{p}} \delta \alpha_{\vec{p}} - \int_\f d^d \vec{k} ~ k_x \Bigl[ |\lambda_{\vec{k}}|^2 + 2 \alpha_{\vec{k}} {\rm Re} \lambda_{\vec{k}}(t)  - \Delta P_\ph(t,\Lambda-\delta \Lambda) 2 \mathcal{M}_{xx}^{-1} k_x \frac{\alpha_{\vec{k}}}{\Omega_{\vec{k}}} \l {\rm Re} \lambda_{\vec{k}}(t) + \alpha_{\vec{k}} \r \Bigr]. \label{eq:tRGfullPph}
\end{flalign}
\end{widetext}	

We start by solving the equation for the phonon momentum. To this end, we note that for any value of the UV cut-off $\Lambda$, we may write for the phonon momentum $P_\ph(t) $ as:
\begin{equation}
P_\ph(t) = P_\ph(t,\Lambda) + \Delta P_\ph(t,\Lambda)  \chi(t,\Lambda).
\label{eq:PphtCleverFlowtRG}
\end{equation}
Before applying the tRG protocol we have
\begin{equation}
P_\ph(t,\Lambda_0) = P_\ph^\MF, \qquad \chi(t,\Lambda_0)=1
\end{equation}
for all times $t$. Although Eq.~\eqref{eq:PphtCleverFlowtRG} is true for arbitrary $\Lambda$, it is not very helpful in most cases because $\Delta P_\ph(t,\Lambda)$ still involves complicated dynamics, see Eq.~\eqref{eq:defDeltaPpht}. However after running the tRG, $ \Delta P_\ph(t,\Lambda \to 0) \to 0$ because there are now phonons left leading to further renormalization when $\Lambda \to 0$. Assuming that $\chi(t,\Lambda)$ does not diverge when $\Lambda \to 0$ we thus obtain
\begin{equation}
P_\ph(t) = \lim_{\Lambda \to 0} P_\ph(t,\Lambda).
\end{equation}
From Eq.~\eqref{eq:tRGfullPph} it is now easy to derive the tRG flow equations \eqref{eq:tRGflowPphLbdat}, \eqref{eq:tRGflowChiLbdat}.

We apply the same trick to calculate the phonon number next. Its most general form at an arbitrary UV cut-off $\Lambda$ is
\begin{equation}
N_\ph(t) = N_\ph(t,\Lambda) + \Delta N_\ph(t,\Lambda) +  \Delta P_\ph(t,\Lambda)  \theta(t,\Lambda).
\end{equation}
Initially, we have for all times $t$ that
\begin{equation}
N_\ph(t,\Lambda_0) = N_\ph^\MF, \qquad \theta(t,\Lambda_0) = 0,
\end{equation}
and the phonon number we want to calculate is given by
\begin{equation}
N_\ph(t) = \lim_{\Lambda \to 0}  N_\ph(t,\Lambda).
\end{equation}
From Eqs.\eqref{eq:tRGfullNph}, \eqref{eq:tRGfullPph} we derive the following tRG flow equations:
\begin{widetext}
\begin{flalign}
\frac{\partial N_\ph(t,\Lambda)}{\partial \Lambda} &= \theta(t,\Lambda) \left\{  2 \int_\s d^d \vec{p} ~p_x  \alpha_{\vec{p}} \frac{ \partial \alpha_{\vec{p}} }{\partial \Lambda} 
- \int_\f d^{d-1}\vec{k} ~ k_x \Bigl[ |\lambda_{\vec{k}}|^2 + 2 \alpha_{\vec{k}} {\rm Re} \lambda_{\vec{k}}(t) \Bigr] \right\}  + \nonumber \\
& \qquad\qquad\qquad\qquad + 2 \int_\s d^d \vec{p} ~  \alpha_{\vec{p}} \frac{ \partial \alpha_{\vec{p}} }{\partial \Lambda} 
- \int_\f d^{d-1}\vec{k} ~  \Bigl[ |\lambda_{\vec{k}}|^2 + 2 \alpha_{\vec{k}} {\rm Re} \lambda_{\vec{k}}(t) \Bigr]  , \\
\frac{\partial \theta(t,\Lambda) }{\partial \Lambda} &= 2 \mathcal{M}_{x x}^{-1} \theta(t,\Lambda)  \int_\f d^{d-1} \vec{k} ~ k_x^2 \frac{\alpha_{\vec{k}}}{\Omega_{\vec{k}}} \l {\rm Re} \lambda_{\vec{k}}(t) + \alpha_{\vec{k}} \r + 2 \mathcal{M}_{x x}^{-1} \int_\f d^{d-1} \vec{k} ~ k_x \frac{\alpha_{\vec{k}}}{\Omega_{\vec{k}}} \l {\rm Re} \lambda_{\vec{k}}(t) + \alpha_{\vec{k}} \r.
\end{flalign}
\end{widetext}

%%%%%%%%%%%%%%%%%%%%%%
\subsection{Time-dependent overlaps and spectral function}
%%%%%%%%%%%%%%%%%%%%%%
Now we turn to the discussion of the time-dependent overlap $A(t)$. The original expression in Eq.~\eqref{eq:defAt} was formulated in the lab frame, but because there are no phonons in the initial state the LLP transformation \eqref{eq:LLPdef} has no effect on this state. Assuming that the non-interacting impurity has a well-defined initial momentum $\vec{P}$ we obtain $A_P(t) = \bra{0} e^{- i  \H_P t} \ket{0}$. 

Next, we introduce the unitary transformation $\hat{U}_\MF$ to change into the frame of quantum fluctuations around the MF solution, obtaining
\begin{align}
A_P(t) &= \bra{0} \hat{U}_\MF e^{- i \hat{U}^\dagger_\MF \H_P \hat{U}_\MF t} \hat{U}^\dagger_\MF \ket{0} \nonumber \\
&=\bra{-\alpha^\MF_{\vec{k}}} e^{- i \tilde{\cal H}(\Lambda_0) t} \ket{-\alpha^\MF_{\vec{k}}}. 
\label{eq:AqtFlucFrameMAIN}
\end{align}
As for the time-dependent observables, we calculate $A_P(t)$ shell-by-shell by applying infinitesimal transformations $\hat{U}_\Lambda$ which diagonalize fast phonons in the Hamiltonian $\tilde{\cal H} = \Ht_\s + \Ht_\sf + \Ht_\f$ in every step. Here $\Ht_\s$ ($\Ht_\f$) contains only slow (fast) phonons and $\Ht_\sf$ defines their coupling. In contrast to the previous cases, the transformations $\hat{U}_\Lambda$ are no longer unitary, although their form is closely related to the unitaries $\hat{U}_\Lambda$ used so far.

%%%%%%%%%%%
\subsubsection{tRG step}
%%%%%%%%%%%
Our starting point is Eq.~\eqref{eq:AqtFlucFrameMAIN}. We start every tRG step by performing the infinitesimal tRG transformation $\hat{W}_\Lambda$,
\begin{equation}
A_P(t) = \bra{- \alpha_{\vec{k}}} \hat{W}_\Lambda e^{- i t \hat{W}_\Lambda^{-1} \tilde{\mathcal{H}} \hat{W}_\Lambda } \hat{W}_\Lambda^{-1} \ket{- \alpha_{\vec{k}}}.
\label{eq:AqtULambdaDerivation}
\end{equation}
In contrast to the previous RG schemes, we only demand that $\hat{W}_\Lambda$ is invertible but it no longer has to be unitary. We will show that during the tRG flow the Hamiltonian $\tilde{\mathcal{H}}$ is no longer hermitian in general. 

To ensure that fast degrees of freedom are diagonalized in the transformed Hamiltonian $\hat{W}_\Lambda^{-1} \tilde{\mathcal{H}} \hat{W}_\Lambda$, we choose $\hat{W}_\Lambda$ to be of the following form,
\begin{equation}
\W_\Lambda =  \exp \l \int_\f d^d \vec{k} ~ \F_{\vec{k}} \left[ \a_{\vec{k}} - \ad_{\vec{k}} \right] \r.
\label{eq:defUdRG}
\end{equation}
This expression is very similar to the unitary transformations used previously, the only difference being that $\a_{\vec{k}}$ is multiplied by $\hat{F}_{\vec{k}}$ instead of $\hat{F}^\dagger_{\vec{k}}$, making the transformation non-unitary in general. As before, we assume that $\hat{F}_{\vec{k}}$ contains slow-phonon operators only. In cases where $\hat{F}_{\vec{k}}^\dagger = \hat{F}_{\vec{k}}$ there is no difference with the unitary case.

Before proceeding with the calculation, let us derive some basic properties of the transformation in Eq.~\eqref{eq:defUdRG}. For simplicity, we will consider a single-mode expression,
\begin{equation}
\hat{\mathcal{D}}_{f} = \exp \left[  f \l \ad - \a \r \right], \qquad f \in \mathbb{C},
\label{eq:DnonUnitaryDef}
\end{equation}
which can be interpreted as a non-unitary generalization of the coherent state displacement operator $\hat{D}_\alpha = \exp \l \alpha \ad -  \alpha^* \a \r$, where $\alpha \in \mathbb{C}$. Using similar manipulations as in the unitary case we can show that
\begin{flalign}
\hat{\mathcal{D}}_{f}^{-1} \a \hat{\mathcal{D}}_{f} & = \a + f, \label{eq:DnonUnitaryProp1} \\
\hat{\mathcal{D}}_{f}^{-1} \ad \hat{\mathcal{D}}_{f} & = \ad + f \label{eq:DnonUnitaryProp2},
\end{flalign}
and, as in the unitary case, $\hat{\mathcal{D}}_f^{-1} = \hat{\mathcal{D}}_{- f}$.

Now we are in a position to generalize the RG protocol to non-hermitian Hamiltonians. To this end ,we apply Eqs.\eqref{eq:DnonUnitaryProp1}, \eqref{eq:DnonUnitaryProp2} and derive an equation for $\F_{\vec{k}}$ in Eq.~\eqref{eq:defUdRG}, such that fast phonons are diagonalized in the new Hamiltonian $\hat{W}_\Lambda^{-1} \tilde{\mathcal{H}} \hat{W}_\Lambda$. Demanding that terms linear in $\ad_{\vec{k}}$ vanish, we obtain
\begin{multline}
 \Omega_{\vec{k}} \F_{\vec{k}} =  \l \alpha_{\vec{k}} - \F_{\vec{k}} \r k_\mu \mathcal{M}_{\mu \nu}^{-1} \int_\s d^d \vec{p} ~  p_\nu \G_{\vec{p}}  \\
 + [\F_{\vec{k}},\H_\s] + \mathcal{O}(\Omega_{\vec{k}}^{-2}),
 \label{eq:selfConsEqFdRG}
\end{multline}
as in the previously discussed RG schemes. For terms linear in $\a_{\vec{k}}$ to vanish, we obtain a separate equation,
\begin{multline}
 \Omega_{\vec{k}} \F_{\vec{k}} =  \l \alpha_{\vec{k}} - \F_{\vec{k}} \r k_\mu \mathcal{M}_{\mu \nu}^{-1} \int_\s d^d \vec{p} ~  p_\nu \G_{\vec{p}}  \\
  - [\F_{\vec{k}},\H_\s] + \mathcal{O}(\Omega_{\vec{k}}^{-2}).
 \label{eq:selfConsEqFdRG2}
\end{multline}

In contrast to the previously discussed RG schemes, the second equation \eqref{eq:selfConsEqFdRG2} poses a condition on $\F_{\vec{k}}$ instead of $\F^\dagger_{\vec{k}}$. The last two equations for $\F_{\vec{k}}$ differ only in a minus sign in front of the commutator $[\F_{\vec{k}},\H_\s]$, which leads to a second order contribution $\mathcal{O}(\Omega_{\vec{k}}^{-2})$. Therefore the leading order solution for $\F_{\vec{k}}$ is the same as in Eq.~\eqref{eq:defFk}, 
\begin{equation}
\F_{\vec{k}} =  \frac{\alpha_{\vec{k}}}{\Omega_{\vec{k}}}  k_\mu \mathcal{M}_{\mu \nu}^{-1} \int_\s d^d \vec{p} ~  p_\nu  \G_{\vec{p}}  + \mathcal{O}(\Omega_{\vec{k}}^{-2}).
\label{eq:FresultdRG}
\end{equation}
As before, the renormalized slow phonon Hamiltonian is of the form 
\begin{equation}
\hat{W}_\Lambda^{-1} \tilde{\mathcal{H}} \hat{W}_\Lambda = \Ht_\s + \delta \Ht_\s + \int d^d \vec{k} ~ \ad_{\vec{k}} \a_{\vec{k}} \l \Omega_{\vec{k}} + \hat{\Omega}_\s(\vec{k}) \r,
\label{eq:HsimplifiedDerivdRGAt}
\end{equation}
see Eqs.\eqref{eq:renHamNonEqMAIN}, \eqref{eq:renHs}. 

Next, we generalize the notion of coherent states to the non-unitary transformations \eqref{eq:DnonUnitaryDef}. To this end, we define $\lambda=(f,\alpha)^T$, $\overline{\lambda}=(f^*,\alpha)^T$ and
\begin{equation}
\ket{\lambda} := \hat{\mathcal{D}}_f \hat{D}_\alpha \ket{0}, \quad \ket{\overline{\lambda}} := \l \hat{\mathcal{D}}_f^{-1} \r^\dagger \hat{D}_\alpha \ket{0} = \hat{\mathcal{D}}_{f^*} \hat{D}_\alpha \ket{0}.
\end{equation}
In addition, we define a positive semidefinite scalar product by
\begin{equation}
\overline{\lambda}^* \lambda := \overline{\lambda}^\dagger \l 
\begin{matrix}
1 & 1 \\
1 & 1
\end{matrix}
\r \lambda = \l f^* + \alpha \r^* \l f + \alpha \r.
\end{equation}
The last equation shows that we may formally set $\lambda = f + \alpha$ and $\overline{\lambda} = f^* + \alpha$ for evaluating the scalar product. 

Before proceeding with the calculation, we derive two more properties of the generalized coherent states. The first concerns the time-dependent overlap for a single mode, for which we find
\begin{equation}
A_\lambda(t) = \bra{\overline{\lambda}} e^{- i \Omega \ad \a t} \ket{\lambda} = \exp \left[  - \l  1 - e^{-i \Omega t} \r \overline{\lambda}^* \lambda \right].
\label{eq:Alambdat}
\end{equation}
The second concerns the time-dependent overlap including the number operator, for which we find the following generalized expression,
\begin{equation}
n_\lambda(t) = \bra{\overline{\lambda}} e^{- i \Omega \ad \a t} \ad \a \ket{\lambda} = A_\lambda(t) \overline{\lambda}^* \lambda e^{- i \Omega t}.
\label{eq:nlambdat}
\end{equation}

We will now evaluate Eq.~\eqref{eq:AqtULambdaDerivation}, but using a more general expression which will be required for the subsequent tRG steps. We make use of the fact that $\alpha_{\vec{k}}^\MF \in \mathbb{R}$ is real, see discussion around Eq.~\eqref{eq:OmegakDef}. In the initial tRG step we can thus write $\ket{- \alpha_{\vec{k}}^\MF}  \equiv \ket{(0,-\alpha_{\vec{k}}^\MF)^T} = \ket{(-\alpha_{\vec{k}}^\MF,0)^T}$. We will now show that the tRG describes a flow of time-dependent overlaps of the form
\begin{equation}
A_P(t) = \bra{(- \alpha_{\vec{k}},0)^T } \hat{W}_\Lambda e^{- i t \hat{W}_\Lambda^{-1} \tilde{\mathcal{H}} \hat{W}_\Lambda } \hat{W}_\Lambda^{-1} \ket{(- \alpha_{\vec{k}},0)^T}.
\label{eq:AqtULambdaDerivation2}
\end{equation}

In order to evaluate Eq.~\eqref{eq:AqtULambdaDerivation2} we need to calculate the action of $\hat{W}_\Lambda^{-1}$ on $\ket{(- \alpha_{\vec{k}},0)^T}$. First we generalize Eq. \eqref{eq:GammaEigenstatesMAIN} and note that $\F_{\vec{k}} \ket{(- \alpha_{\vec{p}},0)^T}_\s = f_{\vec{k}} \ket{(- \alpha_{\vec{p}},0)^T}_\s$, i.e. the slow-phonon generalized coherent MF states $\ket{(-\alpha_{\vec{p}},0)^T}_\s$ are eigenstates of $\F_{\vec{k}}$. Here $f_{\vec{k}}$ is defined as in Eq.~\eqref{eq:def_fkMAIN}. As in the previous tRG scheme for physical observables, we find
\begin{equation}
\hat{W}_\Lambda^{-1} \ket{(-\alpha_{\vec{k}},0)^T}_\f \ket{(- \alpha_{\vec{p}},0)^T}_s = \ket{\lambda_{\vec{k}}}_\f \ket{(- \alpha_{\vec{p}},0)^T}_\s
\label{eq:derivedRGAtLambda1}
\end{equation}
where $\lambda_{\vec{k}}=(f_{\vec{k}}-\alpha_{\vec{k}},0)^T$, and similar
\begin{equation}
~_\s \bra{(- \alpha_{\vec{p}},0)^T} ~_\f \bra{(- \alpha_{\vec{k}},0)^T} \hat{W}_\Lambda = ~_\s \bra{(- \alpha_{\vec{p}},0)^T} ~_\f \bra{ \overline{\lambda}_{\vec{k}}}.
\label{eq:derivedRGAtLambda2}
\end{equation}
Combining Eqs.\eqref{eq:HsimplifiedDerivdRGAt}, \eqref{eq:derivedRGAtLambda1}, \eqref{eq:derivedRGAtLambda2} we can bring the time-dependent overlap Eq.~\eqref{eq:AqtULambdaDerivation2} into the simplified form,
\begin{widetext}
\begin{equation}
A_P(t) =  ~_\s \bra{(-\alpha_{\vec{p}},0)^T}  ~_\f \bra{\overline{\lambda}_{\vec{k}}} e^{- i t \left[ \int_\f d^d \vec{k} ~ \ad_{\vec{k}} \a_{\vec{k}}  \l \Omega_{\vec{k}} + \hat{\Omega}_\s(\vec{k}) \r + \Ht_\s + \delta \Ht_\s \right]}  \ket{\lambda_{\vec{k}}}_\f \ket{(- \alpha_{\vec{p}},0)^T}_\s + \mathcal{O}(\Omega_{\vec{k}}^{-2}).
\end{equation}

As in Sec.~\ref{sec:tRGstep} we proceed by treating the fast phonon frequency renormalization $\propto \hat{\Omega}_\s(\vec{k})$ perturbatively. Using Eqs.\eqref{eq:trotterPert}, \eqref{eq:Alambdat}, \eqref{eq:nlambdat} we obtain
\begin{multline}
A_P(t) =  ~_\s \bra{(-\alpha_{\vec{p}},0)^T} \Biggl[ \Bigl(  \prod_{\vec{k} \in \f}  A_{\lambda_{\vec{k}}}(t) \Bigr)
e^{- i t \l \Ht_\s + \delta \Ht_\s \r} - i \int_0^t d\tau  \int_\f d^d \vec{k} ~  e^{- i \l t - \tau \r \l \Ht_\s + \delta \Ht_\s \r} n_{\lambda_{\vec{k}}}(t) \hat{\Omega}_\s(\vec{k}) e^{- i \tau \l \Ht_\s + \delta \Ht_\s \r} \Biggr] \\ 
\times \ket{(- \alpha_{\vec{p}},0)^T}_\s = \Bigl( \prod_{\vec{k} \in \f} A_{\lambda_{\vec{k}}}(t)  \Bigr) ~_\s \bra{(-\alpha_{\vec{p}},0)^T}  \exp \left[ -i \l \Ht_\s + \delta \Ht_\s + \int_\f d^d \vec{k} ~ \overline{\lambda}_{\vec{k}}^* \lambda_{\vec{k}} e^{- i \Omega_{\vec{k}} t} \hat{\Omega}_\s (\vec{k})  \r t \right] \ket{(- \alpha_{\vec{p}},0)^T}_\s,
\end{multline}
as always up to corrections of order $\mathcal{O}(\Omega_{\vec{k}}^{-2}, \delta \Lambda^2)$. We thus derived the following factorization,
\begin{flalign}
A_P(t) &= A_\f(t) A_\s(t), \label{eq:AqsfMotivationMAIN}  \\
A_\s(t) &= ~_\s \bra{(-\alpha_{\vec{p}},0)^T} \exp \Biggl[ - i \underbrace{\l \Ht_\s + \delta \Ht_\s  + \int_\f d^d \vec{k} ~ \overline{\lambda}_{\vec{k}}^* \lambda_{\vec{k}} e^{-i \Omega_{\vec{k}} t} ~ \hat{\Omega}_\s(\vec{k})  \r }_{=\Ht_\s'}  t \Biggr] \ket{(-\alpha_{\vec{p}},0)^T}_\s, \label{eq:AqtsMotivationMAIN}  \\
A_\f(t) &= \prod_{\vec{k} \in \f} A_{\lambda_{\vec{k}}}(t)  = \exp \left[ - \int_\f d^d \vec{k} ~ \l 1 - e^{- i \Omega_{\vec{k}} t } \r \overline{\lambda}^*_{\vec{k}} \lambda_{\vec{k}} \right].
\label{eq:AqtfMotivationMAIN} 
\end{flalign}
\end{widetext}

Next we will bring the renormalized slow phonon Hamiltonian into a basis of quantum fluctuations around its MF solution. To this end we apply the non-unitary MF shifts 
\begin{equation}
\hat{V}_\MF (\Lambda) = \exp \l \int_\s d^d \vec{p} ~  \delta \alpha_{\vec{p}}  \l \ad_{\vec{p}} - \a_{\vec{p}} \r \r,
\label{eq:VMFdef2}
\end{equation}
which differ from their unitary equivalents \eqref{eq:VMFdef} in that both $\a_{\vec{p}}$ and $\ad_{\vec{p}}$ are multiplied by $\delta \alpha_{\vec{p}}$ in the exponent, cf. Eq.~\eqref{eq:DnonUnitaryDef}. This allows us to deal with the non-unitary Hamiltonian in Eq. \eqref{eq:AqtsMotivationMAIN}.

The calculations to obtain $\delta \alpha_{\vec{p}}$ such that terms linear in $\a_{\vec{p}}$ and $\ad_{\vec{p}}$ vanish in the new Hamiltonian 
\begin{equation}
\Ht(\Lambda-\delta \Lambda) = \hat{V}_\MF^\dagger (\Lambda) \Ht_\s' \hat{V}_\MF(\Lambda)
\end{equation}
are completely analogous to those presented in Sec.~\ref{sec:tRGstep}. They lead to the same form of tRG flow of the generalized coherent amplitudes, given by $\alpha_{\vec{p}} = - V_p / \Omega_{\vec{p}}$. The tRG flow equations for the coupling constants also take a similar form.

%%%%%%%%%%%
\subsubsection{tRG flow equations -- coupling constants}
%%%%%%%%%%%
For the mass $\mathcal{M}_{\mu \nu}$ and the zero-point energy, we obtain the same equations as for time time-dependent observables and as in equilibrium,
\begin{flalign}
\frac{\partial \mathcal{M}_{\mu \nu}^{-1} }{\partial \Lambda} &= 2 \mathcal{M}_{\mu \lambda}^{-1}  \int_\f d^{d-1} \vec{k} ~ \frac{V_k^2}{\Omega^3_{\vec{k}}} k_\lambda k_\sigma  ~\mathcal{M}_{\sigma \nu}^{-1} \label{eq:RGflowMassOverlaps} \\
\frac{\partial E_0}{\partial \Lambda} &= \frac{1}{2} \frac{\partial \mathcal{M}_{\mu \nu}^{-1}}{\partial \Lambda} \int_\s d^d \vec{p} ~ p_\mu p_\nu \l \alpha_{\vec{p}} \r^2.
\label{eq:RGflowZeroPtEnergyDer}
\end{flalign}

For the momentum $\kappa_x$, on the other hand,
\begin{widetext}
\begin{equation}
\frac{\partial \kappa_x }{\partial \Lambda} =- \frac{\partial \mathcal{M}_{xx}^{-1}}{\partial \Lambda} \mathcal{M}_{xx} \kappa_x +  \l 1 + 2 \mathcal{M}_{xx}^{-1} I^{(2)} \r^{-1} 
\left[ 2 \mathcal{M}_{xx}^{-1} I^{(2)}  \l \int_\f d^{d-1}\vec{k} ~ k_x \overline{\lambda}^*_{\vec{k}} \lambda_{\vec{k}} e^{- i \Omega_{\vec{k}} t} \r  - I^{(3)}_{\mu \nu} \frac{\partial \mathcal{M}_{\mu \nu}^{-1} }{\partial \Lambda}   \right] .
\label{eq:MFRGflowkappaXOverlaps}
\end{equation}
The integrals $I^{(2)}$ and $I^{(3)}_{\mu \nu}$ are defined in Eqs.\eqref{eq:I2Def} and \eqref{eq:I3munuDef}. Recall that
\begin{equation}
\lambda_{\vec{k}} = \l -\alpha_{\vec{k}} \left[ 1 + \frac{1}{\Omega_{\vec{k}}}  k_\mu \mathcal{M}_{\mu \nu}^{-1} \int_\s d^d \vec{p} ~  p_\nu  |\alpha_{\vec{p}}|^2   \right] , 0\r^T.
\label{eq:lambdakRecall}
\end{equation}
\end{widetext}

Before discussing the tRG flow equations further, let us point out an important connection to the ground state RG flow. In the long-time limit we can formally set $\lim_{t \to \infty} e^{-i \Omega_{\vec{k}} t} = 0$, i.e. quantum fluctuations have no effect on average. Then comparison to the equilibrium flow equations, see Ref. \cite{Grusdt2016RG}, shows that the tRG is \emph{equivalent} to the ground state flow in the long-time limit.

Although it may be unexpected at first sight that the time-dependent overlap is determined by a non-hermitian Hamiltonian evolution, there is an intuitive explanation why this is the case: Unlike the time-dependent observables $O(t)$ discussed in Sec.~\ref{sec:PhysicalObservalestRG}, the time-dependent overlap describes the amplitude for the state $\ket{0}$ to return to itself after a unitary time evolution $\ket{0} \to e^{- i \Ht t} \ket{0}$. Therefore the information about any contribution which does not return to $\ket{0}$ is completely lost. The corresponding decay of $|A_P(t)|$ in time is described by the imaginary part of the Hamiltonian.

%%%%%%%%%%%
\subsubsection{tRG flow equations -- time-dependent overlap}
%%%%%%%%%%%
In the remainder of this section we derive tRG flow equations for the time-dependent overlap $A_P(t)$, starting from Eqs. \eqref{eq:AqsfMotivationMAIN} - \eqref{eq:AqtfMotivationMAIN}. As $A_P(t)$ factorizes in every RG step, see Eq.~\eqref{eq:AqsfMotivationMAIN}, it is more convenient to consider the logarithm of $A_P(t)$ which (suppressing the index $P$ for simplicity) we denote by
\begin{equation}
B(t) = \log A_P(t).
\end{equation}

Thus, after running the RG from the initial cut-off $\Lambda_0$ down to a value $\Lambda$, we have
\begin{equation}
B(t) = B_\Lambda^>(t) + B_\Lambda^<(t).
\end{equation}
In this expression, the yet unsolved dynamics at smaller momenta is accounted for by
\begin{equation}
B_\Lambda^<(t) = \log \bra{(-\alpha_{\vec{p}},0)^T} e^{- i \Ht(\Lambda) t} \ket{(-\alpha_{\vec{p}},0)^T}, \quad p\leq \Lambda.
\end{equation}
On the other hand, the dynamics at larger momenta are captured by $B_\Lambda^>(t)$, which flows in the RG and starts from
\begin{equation}
B_{\Lambda_0}^>(t)  = 0.
\end{equation}

Let us emphasize again that the time $t$ enters these expressions only as a fixed parameter, while the tRG flow corresponds to a variation of model parameters with $\Lambda$, described by a differential equation of the form $\partial_\Lambda B_\Lambda^>(t) = ...$, see Eq.~\eqref{eq:simplifiedNonEqRGflow} below. At the end of the tRG, we will arrive at a fully converged value for the time-dependent overlap, $B(t) = \lim_{\Lambda \rightarrow 0} B_\Lambda^>(t) + B_\Lambda^<(t)$. While $\lim_{\Lambda \rightarrow 0} B_{\Lambda}^>(t)$ will be determined from a tRG flow equation, $B_\Lambda^<(t)$ contains a $\mathbb{C}$-number contribution $E_0(t)$ flowing in the course of the tRG, plus corrections of order $\mathcal{O}(\Lambda^3)$. Therefore as $\Lambda \rightarrow 0$
\begin{equation}
B(t) = -i E_0(t)  t + \lim_{\Lambda \rightarrow 0} B_\Lambda^>(t),
\label{eq:BtLambdaInf}
\end{equation}
which is the final result of the tRG. As shown above, in the long-time limit the ground state RG flow is recovered, and consequently $E_0(t) \to E_0$ converges to the ground state polaron energy $E_0$ as $t \to \infty$.

With the notations introduced above, we can now proceed by discussing the tRG flow equations for $B(t)$. For a single tRG step we obtain from Eqs.\eqref{eq:AqsfMotivationMAIN} - \eqref{eq:AqtsMotivationMAIN}
\begin{equation}
B(t) = B_\Lambda^>(t) + \delta B_\Lambda + B_{\Lambda-\delta \Lambda}^<(t),
\end{equation}
where we read off (in generalization of the usual coherent states)
\begin{flalign}
\delta B_\Lambda &= \log ~_\f \bra{\overline{\lambda}_{\vec{k}}}  e^{- i t \int_\f d^d \vec{k} ~ \Omega_{\vec{k}} \ad_{\vec{k}} \a_{\vec{k}} } \ket{\lambda_{\vec{k}}}_\f \nonumber \\
&= - \int_\f d^d \vec{k} ~ \overline{\lambda}_{\vec{k}}^* \lambda_{\vec{k}} \l 1 - e^{-i \Omega_{\vec{k}} t} \r.  \label{eq:AfContribution}
\end{flalign}
Thus, we arrive at the following tRG flow equation,
\begin{equation}
\frac{\partial B^>_\Lambda(t)}{\partial \Lambda} = \int_\f d^{d-1} \vec{k} ~  \overline{\lambda}_{\vec{k}}^* \lambda_{\vec{k}} \l 1 - e^{-i \Omega_{\vec{k}} t} \r.
\label{eq:simplifiedNonEqRGflow}
\end{equation}

Some comments are in order about the tRG flow equation \eqref{eq:simplifiedNonEqRGflow}. To begin with, we note that in the limit $t \rightarrow \infty$ the complex phase factors $e^{- i \Omega_{\vec{k}} t}$ vanish because of dephasing, and we can effectively set $\lim_{t \rightarrow \infty} e^{- i \Omega_{\vec{k}} t} = 0$. Thus, by comparing to the RG flow equation of the logarithm of the quasiparticle weight $\log Z$, see Ref.~\cite{Grusdt2015Varenna}, and employing Eq.~\eqref{eq:BtLambdaInf}, we obtain
\begin{flalign}
\lim_{t \rightarrow \infty} B(t) & \equiv \lim_{t \rightarrow \infty} \log A_P(t) \nonumber \\
&= - i E_0 t - \log Z. 
\label{eq:Btasympt}
\end{flalign}
This result represents an important consistency check for the tRG procedure; it can be shown rigorously, using a standard Lehman-type spectral decomposition, that in the long-time limit the time-dependent overlap $A_P(t)$ is determined solely by ground state properties (see Ref. \cite{Shashi2014RF} for a discussion).

A second important remark concerns the relation of the tRG flow equation \eqref{eq:simplifiedNonEqRGflow} to the MF result for the time-dependent overlap $A_P(t)$. In Appendix \ref{apdx:MFAPt}, we discuss the spherically symmetric situation (i.e. $P=0$) and show that both expressions for $A_P(t)$ (from tRG and MF) have an almost identical form in that case. To obtain the MF expression one has to merely discard the tRG flow, i.e. replace $\Omega_{\vec{k}} \rightarrow \Omega_{\vec{k}}^\MF$ and $\lambda_{\vec{k}}, \overline{\lambda}_{\vec{k}} \rightarrow - \alpha^\MF_{\vec{k}}$ in the tRG expression, and drop energy corrections $\Delta E$ in the expression for $E_0$ due to the RG.

%%%%%%%%%%%%%%%%%%%%%%%%%%%%%%%%%%%%%%%%%%%%%%%%%%%%%
\section{Summary and Outlook}
\label{sec:Summary}
%%%%%%%%%%%%%%%%%%%%%%%%%%%%%%%%%%%%%%%%%%%%%%%%%%%%%
We developed a time-dependent renormalization group (tRG) technique to solve far-from-equilibrium quantum impurity problems. We applied the method to the ubiquitous class of Fr\"ohlich Hamiltonians, for which we presented derivations of the tRG flow equations. We demonstrated that our approach allows calculation of the spectral function as well as the formation dynamics of polarons. We analyzed the latter by studying impurity trajectories following an interaction quench.

We applied the tRG method to analyze the dynamics of impurity atoms inside a Bose-Einstein condensate. At weak couplings, the Fr\"ohlich Hamiltonian provides an accurate description of this problem. We also studied corrections beyond the Fr\"ohlich paradigm which need to be included at stronger couplings. For the spectral function, we generalized the tRG equations to include two-phonon terms when the total polaron momentum is zero. We also performed time-dependent mean-field calculations of the full model within the Bogoliubov approximation and calculated polaron trajectories. On the attractive side of a Feshbach resonance, we demonstrated that the main effect of the additional two-phonon terms is to renormalize the coupling constant of the effective Fr\"ohlich model, even far from equilibrium. Therefore we expect that our predictions are relevant for current experiments with ultracold atoms. 

For light impurities in the intermediate coupling regime, we predict non-trivial polaron trajectories following a sudden interaction quench. We have shown that the impurity can be dramatically slowed down, which can serve as an indirect indicator for strong polaronic mass renormalization in the system. Currently, it is unclear how large the effective polaron mass is for the experimentally realized Bose polarons at strong couplings \cite{Jorgensen2016PRL,Hu2016PRL}. While experimental measurements are still lacking, the question has been controversially discussed by theorists \cite{Rath2013,Jorgensen2016PRL,Grusdt2017RG}. We expect, therefore, that experiments realizing polaron dynamics as discussed in this paper can shed new light on this question. 

In the spectral function, we found a dramatic shift of spectral weight to higher energies in the strong coupling regime for light impurities. This is consistent with recent measurements \cite{Jorgensen2016PRL,Grusdt2017RG}.

Our theoretical method can also be applied to even stronger couplings. Our initial calculations showed that the polaron trajectories could reverse direction after the interaction quench for light impurities and strong couplings. However, it remains unclear how reliable this prediction is, and this regime will be explored further in future work.

We now comment on possible extensions of our work beyond the polaron problems. The basic idea of our approach is to diagonalize phonon modes step by step at different momenta. Each mode is described by a free harmonic oscillator in the new frame of reference chosen during the tRG protocol. In general, this type of methodology is well suited for solving problems involving multiple timescales. It can be easily generalized to analyze other far-from-equilibrium problems. A system closely related to the quantum impurity problem is the so-called angulon, i.e. a problem of rotational excitations of molecules immersed in a quantum fluid \cite{Schmidt2015,Schmidt2016PRX,Lemeshko2017,Bighin2017}. This system was shown to exhibit quasiparticles similar to polarons, but with a conserved total angular momentum and a cloud of angular excitations renormalizing the moments of inertia.  We expect that the tRG approach can be straightforwardly generalized to study far-from-equilibrium dynamics of angulons. 

The tRG method can be generalized to problems with an explicit time-dependence, as demonstrated in Appendix \ref{apdx:fullTimeDeptRG}. This allows calculating different types of dynamics relevant in ongoing experiments with ultracold atoms, including averaging over trap effects, finite ramping times through Feshbach resonances, or polaron transport in the presence of an external force. Another important aspect of cold atom experiments are effects of finite temperatures \cite{Guenther2017,Levinsen2017,Lausch2017,Pastukhov2017}. By keeping track of thermal populations of phonon modes in the tRG, we expect that our approach can also be generalized to address far from equilibrium polaron problems at finite temperatures in the future.

%%%%%%%%%%%%%%%%%%%%%%%%%%%%%%%%%%%%%%%%%%%%%%%%%%%%%
\section*{Acknowledgements}
%%%%%%%%%%%%%%%%%%%%%%%%%%%%%%%%%%%%%%%%%%%%%%%%%%%%%
We acknowledge fruitful discussions with R. Schmidt, A. Shashi, T. Lausch, C. Gohle, I. Bloch, D. Pekker, D. Abanin, A. Widera, I. Cirac, M. Banuls, A. Rubtsov, M. Knap, M. Lemeshko and G. Bighin. Support from Harvard-MIT CUA, NSF Grant No. DMR-1308435 and from AFOSR Quantum Simulation MURI, AFOSR grant number FA9550-16-1-0323 is gratefully acknowledged. F.G. acknowledges support from the Gordon and Betty Moore Foundation under the EPIQS program, the Marion K\"oser Stiftung and the physics department of the TU Kaiserslautern. K.S. acknowledges financial support from the Department of Defense NDSEG Fellowship. 

\appendix
%\newpage

%%%%%%%%%%%%%%%%%%%%%%%%%%%%%%%%%%%%%%%%%%%%%%%%%%%%%
\section{tRG for Hamiltonians with explicit time-dependence}
\label{apdx:fullTimeDeptRG}
%%%%%%%%%%%%%%%%%%%%%%%%%%%%%%%%%%%%%%%%%%%%%%%%%%%%%
In this appendix we generalize our tRG method to Hamiltonians $\H(t)$ with an explicit time-dependence. This situation appears naturally, for example, when an external force $\vec{F}$ is applied to the impurity and the total system momentum $\vec{P}(t) = \vec{P}(0) + \int_0^t d\tau \vec{F}(\tau)$ in the LLP Hamiltonian becomes time-dependent. Another example is when interactions are ramped up slowly instead of considering an infinitesimal interaction quench as above. This scenario is naturally realized when a magnetic field is ramped through a Feshbach resonance in a finite time, making the coupling constant $\alpha(t)$ time-dependent. 

%%%%%%%%%%%%%%%%%%%%%%%%%%%
\subsection{Derivation}
%%%%%%%%%%%%%%%%%%%%%%%%%%%
Our goal is to calculate time-dependent observables such as the phonon number. As before, the idea is to expand around the instantaneous MF solution $\alpha_{\vec{k}}(t)$ of the Hamiltonian $\H_{\vec{P}}(t)$. Starting from the phonon vacuum $\ket{\psi_0}=\ket{0}$ as in the main text, we obtain $N_\ph(t) = N_\ph^\MF(t) + \Delta N_\ph(t)$, where $N_\ph^\MF(t)$ is the MF phonon number obtained from the instantaneous MF solution $\alpha_{\vec{k}}(t)$ of the Hamiltonian $\H_{\vec{P}}(t)$. Corrections are given by
\begin{widetext}
\begin{equation}
 \Delta N_\ph(t) = \int^{\Lambda_0} d^d \vec{k} ~ \bra{-\alpha_{\vec{\kappa}}^\MF(0)}    \mathcal{T} e^{i \int_0^t d\tau~ \tilde{\mathcal{H}}_\eff (\tau) }     \G_{\vec{k}}(\Lambda_0,t)   \mathcal{T}  e^{- i \int_0^t d\tau~ \tilde{\mathcal{H}}_\eff(\tau) }    \ket{- \alpha_{\vec{\kappa}}^\MF(0)}.
 \label{eq:DeltaNphtRGHt}
\end{equation}
\end{widetext}
Here the effective Hamiltonian is given by
\begin{equation}
\tilde{\mathcal{H}}_\eff(t) = \tilde{\mathcal{H}}(t)  + i \int^{\Lambda_0} d^d \vec{k} ~ \left[ \l \frac{\partial \alpha^*_{\vec{k}}(t)}{\partial t} \r \a_{\vec{k}} - \hc \right],
\label{eq:nonAdiabCorr}
\end{equation}
where the first term is the Hamiltonian \eqref{eq:HquantFluc2} obtained from expanding around the instantaneous MF solution $\alpha_{\vec{k}}(t)$, and the second term describes non-adiabatic corrections. 

To apply the same diagonalization of fast phonons as in the main text, we introduce a Trotter decomposition in Eq.~\eqref{eq:DeltaNphtRGHt}. This leaves us with a discrete set of time steps $\tau_1,...,\tau_N$ with a spacing $\delta \tau$. At any infinitesimal step, we can diagonalize fast phonons by inserting the unitary transformations $\hat{W}_\Lambda(\tau_j)$, see Eq.~\eqref{eq:WMF-RG}, defined now for the instantaneous Hamiltonians $\tilde{\mathcal{H}}(\tau_j)$. First of all, this leads to additional non-adiabatic corrections, because
\begin{equation}
\hat{W}_\Lambda^\dagger(\tau_{j+1}) \hat{W}_\Lambda(t_j) = \exp \left[ \delta \tau  \int_\f d^d \vec{k} ~ \ad_{\vec{k}} \frac{\partial \F_{\vec{k}}(\tau_j)}{\partial t}  - \hc \right].
\end{equation}
As these terms go like $\F_{\vec{k}} = \mathcal{O}(\Omega_{\vec{k}}^{-1})$, they can be neglected and we obtain the same solution $\F_{\vec{k}}(t)$ required for the diagonalization of fast phonons, Eq.~\eqref{eq:defFk} as before.

Next, we calculate the renormalized Hamiltonian, $\hat{W}_\Lambda^\dagger(\tau) \tilde{\mathcal{H}}_\eff(\tau) \hat{W}_\Lambda(\tau)$. The non-adiabatic corrections in Eq.~\eqref{eq:nonAdiabCorr} lead to an additional renormalization of the slow-phonon Hamiltonian, 
\begin{equation}
 i \int_\f d^d \vec{k} \left[ \F^\dagger_{\vec{k}} \frac{\partial \alpha_{\vec{k}}}{\partial t} - \F_{\vec{k}} \frac{\partial \alpha^*_{\vec{k}}}{\partial t}  \right],
\end{equation}
which vanishes when $\alpha_{\vec{k}}(t) \in \mathbb{R}$ is real (such that $\F^\dagger_{\vec{k}}=\F_{\vec{k}}$). As we expand around the instantaneous MF solutions, we have enough gauge freedom to make this choice.

The most important difference to the time-independent protocol is that the fast phonon Hamiltonian now contains non-adiabatic corrections linear in $\a_{\vec{k}}$, see Eq.~\eqref{eq:nonAdiabCorr}. Hence, the dynamics of the fast-phonon coherent states $\ket{\lambda_{\vec{k}}(\tau)}$, cf. Eq.~\eqref{eq:lambdakt}, are determined by	
\begin{equation}
i \partial_t \ket{\lambda_{\vec{k}}(t)} = \left[ \Omega_{\vec{k}}(t) \ad_{\vec{k}} \a_{\vec{k}} + i \frac{\partial \alpha_{\vec{k}}(t)}{\partial t} \l \a_{\vec{k}} - \ad_{\vec{k}} \r \right] \ket{\lambda_{\vec{k}}(t)}.
\label{eq:lamdaktHt}
\end{equation}
The initial condition is the same as before, 
\begin{equation}
\lambda_{\vec{k}}(0) = - \alpha_{\vec{k}}(0) \left[ 1 + \frac{k_\mu \mathcal{M}_{\mu \nu}^{-1}(t=0)}{\Omega_{\vec{k}}} \int_\s d^d \vec{p} ~ p_\nu \alpha_{\vec{p}}^2(0) \right].
\label{eq:lamdak0Ht}
\end{equation}

Finally, we can introduce the tRG flow of the MF amplitudes $\alpha_{\vec{k}}(\Lambda,t)$ by inserting unitary transformation $\hat{V}_\MF(\Lambda,\tau_j)$ as in Eq.~\eqref{eq:VMFdef} in the Trotter decomposition. This changes the Hamiltonians $\tilde{\mathcal{H}}(\tau_j)$ as before. The non-adiabatic corrections \eqref{eq:nonAdiabCorr} retain their form, but the MF amplitudes $\alpha_{\vec{k}}(\Lambda,t)$ depend on both $\Lambda$ and $t$.

%%%%%%%%%%%%%%%%%%%%%%%%%%%
\subsection{Result}
%%%%%%%%%%%%%%%%%%%%%%%%%%%
Summarizing the results from above, we end up with the following universal Hamiltonian,
\begin{multline}
\tilde{\mathcal{H}}(\Lambda,t)= E_0(\Lambda,t) + \int^{\Lambda} d^d \vec{k} ~ \ad_{\vec{k}} \a_{\vec{k}} \Omega_{\vec{k}}(\Lambda,t)  \\
+ \int^{\Lambda} d^d \vec{k} ~ d^d \vec{k}' ~ \frac{k_\mu \mathcal{M}_{\mu \nu}^{-1}(\Lambda,t) k_\nu' }{2}  : \G_{\vec{k}}(\Lambda,t) \G_{\vec{k}'}(\Lambda,t) : \\
+ i  \int^{\Lambda} d^d \vec{k} ~ \frac{\partial \alpha_{\vec{k}}(\Lambda,t)}{\partial t} \l \a_{\vec{k}} - \ad_{\vec{k}} \r.
\label{eq:HquantFlucHt}
\end{multline}
similar to Eq. \eqref{eq:HquantFluc2}.

For every point in time, $t$, we obtain the same tRG flow equations as in Eqs.~\eqref{eq:RGflowMass}, \eqref{eq:MFRGflowkappaX}, \eqref{eq:RGflowZeroPtEnergy} - \eqref{eq:tRGflowChiLbdat}. Note that the coupling constants $\mathcal{M}_{\mu \nu}^{-1}(\Lambda,t)$ and $\kappa_x(\Lambda,t)$ as well as $E_0(\Lambda,t)$, become explicitly time-dependent. Most importantly, $\lambda_{\vec{k}}(t)$ is now determined by Eqs.\eqref{eq:lamdaktHt}, \eqref{eq:lamdak0Ht}. These equations contain information about all higher momenta and all previous times. Nevertheless they are easy to solve numerically, because different $\vec{k}$-modes can be considered independently.

%%%%%%%%%%%%%%%%%%%%%%%%%%%%%%%%%%%%%%%%%%%%%%%%%%%%%
\section{Mean-field expression for time-dependent overlaps}
\label{apdx:MFAPt}
%%%%%%%%%%%%%%%%%%%%%%%%%%%%%%%%%%%%%%%%%%%%%%%%%%%%%
As discussed in Ref.~\cite{Shashi2014RF}, the time-dependent overlap $A_P(t)$ can easily be calculated from a time-dependent variational MF wavefunction, $\ket{\psi(t)} = e^{- i \chi(t)} \prod_{\vec{k}} \ket{\alpha_{\vec{k}}(t)}$. In this appendix we clarify the relation between the time-dependent overlap $A_P(t)$ obtained from the MF variational calculation and from the dynamical RG protocol. For simplicity we focus on the spherically symmetric case when the total system momentum $P=0$ vanishes. 

We start by deriving an exact analytic expression for the time-dependent overlap $A_0^\MF(t)$ obtained from the variational MF ansatz. To this end, we solve the equations of motion for $\alpha_{\vec{k}}(t)$ and $\chi(t)$, obtained from Dirac's time-dependent variational principle in Ref.~\cite{Shashi2014RF}. The initial condition is $\alpha_{\vec{k}}(0)=0$, which respects the spherical symmetry. Because $P=0$ we see that $\vec{P}^\MF_\ph(t)=0$ for all times $t$. This leads to the exact expressions
\begin{flalign}
\alpha_{\vec{k}}(t) &= - i V_k \int_0^t d\tau ~ e^{- i \Omega^\MF_{\vec{k}} \l t - \tau \r }, \label{eq:alphaMFtexact}\\
\chi(t) &= {\rm Re} \int_0^t d \tau \int d^d \vec{k} ~ V_k \alpha_{\vec{k}}(\tau), \label{eq:chiMFtexact}
\end{flalign}
where $\Omega^\MF_{\vec{k}} = \omega_k + \frac{k^2}{2 M}$.

From the variational wavefunction one directly derives the following result for the time-dependent overlaps,
\begin{equation}
A^\MF_0(t) = e^{- i \chi(t)- \frac{1}{2} N_\ph^\MF(t)}, \quad N_\ph^\MF(t) = \int d^d \vec{k} ~ |\alpha_{\vec{k}}(t)|.
\end{equation}
Using Eqs. \eqref{eq:alphaMFtexact} and \eqref{eq:chiMFtexact} in this expression, we end up with an analytical expression for $A^\MF_0(t)$,
\begin{equation}
A_0^\MF(t) = e^{\int d^d \vec{k} ~ \frac{V_k^2}{\Omega^\MF_{\vec{k}}} \left[  i t - \frac{1}{\Omega_{\vec{k}}^\MF}  \l 1 - e^{- i \Omega_{\vec{k}}^\MF t } \r \right] }.
\label{eq:A0MFtapdx}
\end{equation}

The variational MF theory becomes integrable in the spherically symmetric case because of the absence of coupling elements between phonons of different momenta $\vec{k} \neq \vec{k}'$. When quantum fluctuations are included, we expect such couplings to become relevant, and this is indeed the case for the Hamiltonian in Eq.~\eqref{eq:HquantFluc2}. The goal of the dynamical RG, on the other hand, is to eliminate these couplings by a series of infinitesimal unitary transformations, which we achieve perturbatively in every momentum shell. We end up with a simpler but renormalized Hamiltonian for every momentum shell, the dynamics of which is solved exactly by a MF wavefunction.

Indeed, integrating the tRG flow equations \eqref{eq:simplifiedNonEqRGflow} for $A_P(t)$ in the case $P=0$, we obtain
\begin{equation}
A_0^\text{RG}(t) = e^{  - i t \Delta E  + i t \int d^d \vec{k} ~ \frac{V_k^2}{\Omega_{\vec{k}}}   - \int d^d \vec{k} ~ \overline{\lambda}^*_{\vec{k}}\lambda_{\vec{k}} \l 1 - e^{- i \Omega_{\vec{k}} t } \r }.
\label{eq:A0RGtapdx}
\end{equation}
Here, $\Delta E = E_0 - E_0^\MF$ denotes polaron ground state energy corrections by the RG and $\Omega_{\vec{k}}$ is the renormalized phonon frequency, see Eq.~\eqref{eq:OmegaDefRG}, where the coupling constant $\mathcal{M}_{\mu \nu}^{-1}(\Lambda)$ has to be evaluated at $\Lambda=k$ (note that $\kappa_\mu=0$ when $P=0$). Moreover, $\lambda_{\vec{k}}$ is defined in Eq.~\eqref{eq:lambdakRecall}, and to leading order
\begin{equation}
\lambda_{\vec{k}} = - \alpha_{\vec{k}} \l 1 + \mathcal{O}(\Omega_{\vec{k}}^{-1})  \r.
\end{equation}

In conclusion, the form of Eq.~\eqref{eq:A0RGtapdx} is closely related to the MF result Eq.~\eqref{eq:A0MFtapdx}. If the tRG flow is discarded, the MF result is exactly reproduced by the tRG approach.

%%%%%%%%%%%%%%%%%%%%%%%%%%%%%%%%%%%%%%%%
%\bibliography{/Users/fgrusdt/Documents/Library/dataBase_JabRef2.bib}
%\bibliographystyle{unsrt}

\end{document}